\documentclass{article}
\PassOptionsToPackage{numbers, compress}{natbib}
\usepackage[table]{xcolor}
\usepackage[preprint]{neurips_2026}
\usepackage[utf8]{inputenc}
\usepackage[T1]{fontenc}
\usepackage{hyperref}
\usepackage{url}
\usepackage{booktabs}
\usepackage{amsfonts}
\usepackage{amsmath}
\usepackage{amssymb}
\usepackage{nicefrac}
\usepackage{microtype}
\usepackage{xcolor}
\usepackage{graphicx}
\usepackage{multirow}
\usepackage{enumitem}
\usepackage{subcaption}
\usepackage{mathtools}
\usepackage{array}
\newcolumntype{L}[1]{>{\raggedright\arraybackslash}p{#1}}

\title{Beyond Prediction Accuracy: Target-Space Recovery Profiles for Evaluating Model--Brain Alignment}

\author{
Ken Nakamura$^{1}$\,
Tomoya Nakai$^{1}$\,
Ryuto Yashiro$^{1,2}$\,
Ayumu Yamashita$^{3}$\,
Kaoru Amano$^{1}$\\[0.6em]
\begin{tabular}{c}
$^{1}$The University of Tokyo\\
$^{2}$Osnabr\"uck University and Freie Universit\"at Berlin\\
$^{3}$Kobe University
\end{tabular}\\[0.6em]
\texttt{nakamuraken1007@g.ecc.u-tokyo.ac.jp}
}

\newcommand{\RR}{\mathbb{R}}
\newcommand{\tr}{\operatorname{tr}}
\newcommand{\med}{\operatorname{med}}
\newcommand{\TopKCov}{\operatorname{TopKCov}}
\newcommand{\ProfileMean}{\operatorname{ProfileMean}}
\newcommand{\BrainRefScore}{\operatorname{BrainRefScore}}
\newcommand{\DirCov}{\operatorname{DirCov}}
\newcommand{\TargetRef}{\operatorname{TargetRef}}
\newcommand{\FullRefCov}{\operatorname{FullRefCov}}

\begin{document}
\maketitle

\begin{abstract}
Artificial vision models are often evaluated against the human visual cortex by measuring how accurately their internal representations predict brain responses. However, prediction accuracy alone does not indicate which dimensions of the target brain's response space are recovered. Here, we introduce a unified framework for evaluating both model--brain and brain--brain alignment by identifying the response dimensions recovered by prediction. Using repeated fMRI measurements, we first identify target-brain response dimensions that can be reproducibly predicted across independent trial splits. We then predict target-brain responses from either another subject's brain responses or a vision model's internal representations, and quantify how strongly each of these reproducible response dimensions is recovered. Applying this framework to a subset of the Natural Scenes Dataset, in which eight subjects viewed the same natural images during fMRI, we find that the early-to-intermediate visual-cortex responses contain a low-dimensional set of reproducible dimensions. Brain-to-brain comparisons identify which of these dimensions are consistently recoverable from other subjects' brains, providing a diagnostic human reference rather than only a scalar benchmark. In some cases, pretrained and randomly initialized models achieve similar prediction accuracy while showing distinct recovery profiles across these response dimensions. These results show that prediction accuracy alone can mask model--brain mismatches. By making explicit which reproducible brain response dimensions are recovered by prediction, our framework provides a more diagnostic evaluation of alignment between artificial vision models and the human visual cortex.
\end{abstract}

\section{Introduction}

One of the central goals of NeuroAI is to use computational models to understand mechanisms of brain processing and to guide the design of artificial systems \citep{yamins2016goaldriven,zador2023catalyzing}. For visual processing, this goal is often pursued by comparing how artificial models and the human visual cortex respond to the same images, especially using deep vision models as predictors of visual cortical responses \citep{yamins2014performance,guclu2015deep}. Encoding analyses evaluate how accurately model representations predict brain responses \citep{naselaris2011encoding}, whereas representational analyses compare the geometry of model and brain response patterns over a shared stimulus set \citep{kriegeskorte2008rsa,kornblith2019cka}. Recent large-scale fMRI datasets provide repeated measurements, shared stimuli across subjects, single-trial response estimates, and region-of-interest (ROI)-resolved responses in the visual cortex, making it possible to compare not only models with human brains but also different human brains with one another \citep {allen2022nsd,prince2022glmsingle,gifford2026nsdsynthetic}.

\paragraph{Prediction accuracy does not reveal which response dimensions are recovered.}

Model--brain comparisons often summarize alignment by a single score: how accurately model representations predict brain responses. This score is useful for comparing models, but it does not indicate which dimensions of the target brain's response space are recovered by the prediction. For a fixed ROI, the target response space is the voxel or vertex space of that ROI: each image gives one response vector in this space, defined by its response pattern across voxels or vertices. A prediction can match target responses well on average while recovering only some directions in this space, or while recovering directions that differ from those recovered by another prediction with similar accuracy.

Repeated measurements allow us to identify target-brain response dimensions that can be reproducibly predicted across independent trial splits. These dimensions are not defined by response variance alone. Instead, they are defined by predictive reproducibility within the target brain: dimensions that are repeatedly recovered when one split of the target-brain responses is predicted from another. We call the resulting target-brain reference the \emph{reproducible target reference}, and refer to its dimensions as \emph{target-reference dimensions}. Informally, a target-reference dimension is recovered by a source when the target responses predicted from the source contain that dimension. To understand model--brain alignment more diagnostically, we therefore need to ask not only how accurately a model predicts brain responses, but also how much the predicted responses overlap with these reproducible dimensions and which target-reference dimensions are recovered.

\paragraph{Brain-to-brain prediction provides a human reference.}

This question is especially important because measured brain responses contain both reproducible signals and noise. Moreover, some target-reference dimensions may be consistently recoverable from other subjects' brains, whereas others may be more specific to an individual brain. Thus, before asking whether a vision model is brain-like, we first ask which response dimensions of the target brain can be recovered from another subject viewing the same images.

Brain-to-brain prediction provides this human reference. For each target brain, we use other subjects' brain responses to predict its responses. This comparison is not used only as a scalar benchmark or ceiling-like reference, as in conventional noise-ceiling analyses \citep{lage2019methods,schoppe2016performance}. Instead, it identifies which target-reference dimensions are typically recoverable from other human brains. A model can then be evaluated against the same brain-to-brain human reference: does it recover the same target-reference dimensions as other human brains, or does it achieve similar prediction accuracy while recovering different target-reference dimensions? This use of other brains as a reference is related to intersubject-response analyses, which use an individual's responses to quantify shared information across brains \citep{nastase2019sharedresponses}.

\paragraph{Our approach: compare which response dimensions each prediction recovers.}

Here, we introduce a framework for evaluating both model--brain and brain--brain alignment by asking which dimensions of the target brain's response space are recovered by prediction. We use the term \emph{source} to refer to the input side of a source-to-target prediction: either another subject's brain responses or a vision model's internal representations. For each target brain, we first use repeated fMRI measurements to estimate the reproducible target reference, that is, the target-brain response dimensions that can be reproducibly predicted across independent trial splits. We then use each source to predict target responses in that same response space. The target response patterns predicted from a source span a \emph{predictive subspace}: the target-space subspace induced by that source-to-target prediction.

The predictive subspace need not match the reproducible target reference. It may overlap strongly with the reference, overlap only weakly with it, or overlap with a different set of target-reference dimensions than another source. Our analysis, therefore, asks two questions. First, how much do the predicted responses overlap with the reproducible target reference? Second, within that reference, which target-reference dimensions are recovered? Answering these questions enables us to distinguish sources that achieve similar prediction accuracy yet recover different response dimensions within the target brain's response space.

\paragraph{Relation to prior work.}

Our work builds on two lines of research.

First, several studies have shown that the scalar model--brain scores can hide important structure. Similar prediction accuracy can arise from different spectral structures of the model representation and different alignments between those structures and brain responses \citep{canatar2023spectral}; model dimensions that contribute to brain prediction are not always recoverable from brain activity \citep{muzellec2026reverse}; and models with similar ROI-wise scores can differ in whether they reproduce brain-like relationships across regions \citep{hoefling2026onlybrains}. Together, these studies show that prediction accuracy alone is insufficient to explain model--brain alignment.

Second, work on functional alignment and neural population interactions shows that brain responses contain structured components that can be shared or selectively involved in prediction. Hyperalignment and a shared response model learn common spaces that align multi-subject fMRI responses despite individual differences \citep{haxby2011hyperalignment,chen2015srm}. Communication subspace analyses show that prediction between neural populations can depend on a selective low-dimensional subset of activity patterns \citep{semedo2019communication}. These studies motivate the question of which response dimensions are shared or recoverable across systems. 

Our goal is to bring these ideas into a common evaluation framework for both model--brain and brain--brain comparison. We do not decompose scores in the model's coordinate system, test the recoverability of model units, compare cross-region relationships in the brain, or learn a new common space across subjects. Instead, we fix the reproducible target reference in each target brain's response space using repeated measurements. We then ask how much the predicted responses from each source overlap with this reference, and which target-reference dimensions are recovered.

\paragraph{Contributions.}
This study makes three contributions:
\begin{enumerate}
    \item We introduce a target-space framework for evaluating both model--brain and brain--brain alignment by identifying which response dimensions of a target brain are recovered by predictions from models or other human brains.
    \item We use repeated fMRI measurements to identify target-brain response dimensions that can be reproducibly predicted across independent trial splits, and use brain-to-brain prediction to characterize which of these dimensions can be recovered from other human brains.
    \item Using a subset of the Natural Scenes Dataset, we show that models with nearly identical prediction accuracy can differ in the degree and pattern of recovery across these reproducible response dimensions.
\end{enumerate}
The results show that prediction accuracy alone can mask model--brain mismatches. Our framework reveals these mismatches by making explicit which reproducible response dimensions each prediction recovers.

\section{Predictive subspace framework}
\label{sec:framework}
\paragraph{Definitions and overview.}
A \emph{source} is the input side of a source-to-target prediction: either another subject's brain responses or a vision model's internal representations. For source $s$, let $X_s\in\RR^{n\times p_s}$ denote its data matrix, where $n$ is the number of images and $p_s$ is the dimensionality of the source representation. A \emph{target unit} is a subject--hemisphere--ROI combination, and its \emph{target response space} is the voxel/vertex response space of that unit. For a target unit, let $Y\in\RR^{n\times q}$ denote the target response matrix, where $n$ is the number of images and $q$ is the number of retained voxels or vertices. Repeated target-brain responses define the \emph{reproducible target reference}, whose dimensions are \emph{target-reference dimensions}. A source-to-target fit spans a \emph{source-induced predictive subspace} in the same target response space.

\begin{figure*}[h]
    \centering
    \includegraphics[width=\linewidth]{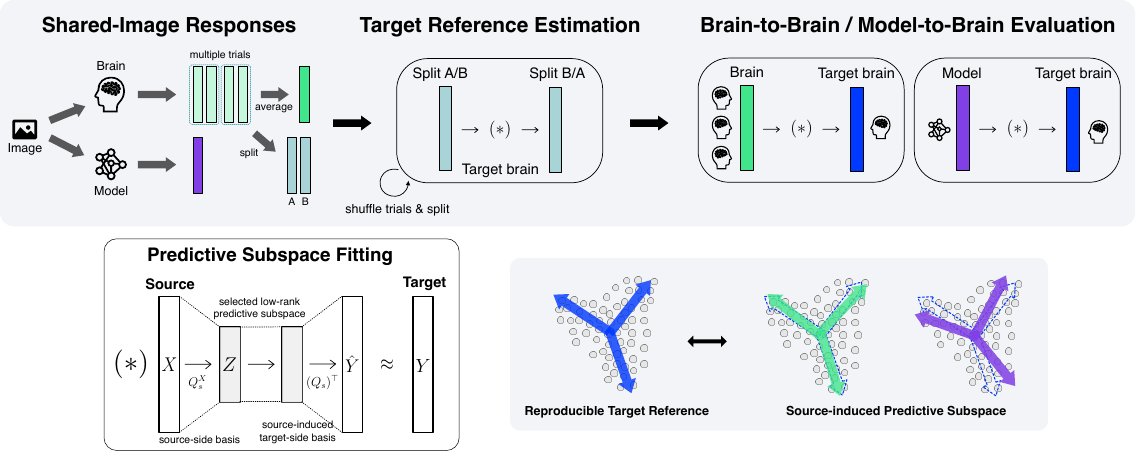}
    \caption{\textbf{Overview of the predictive subspace framework.} Repeated target-brain responses are split into two averaged views, and the predictive subspace fitting operator $(*)$ between views estimates the \emph{reproducible target reference}. The same operator is applied to each source, either another subject's brain responses or a model representation, to predict target responses. The predicted response patterns span a \emph{source-induced predictive subspace}. Comparing this subspace with the reproducible target reference shows which target-reference dimensions are recovered.}
    \label{fig:framework}
\end{figure*}

The framework has two steps (Figure~\ref{fig:framework}).

First, repeated target-brain responses define the reproducible target reference by identifying dimensions recovered across independent trial splits.

Second, each source predicts target responses, yielding a source-induced predictive subspace. Prediction accuracy measures aggregate prediction quality; the structural analysis asks how strongly this subspace overlaps with the reproducible target reference and which target-reference dimensions are recovered.

\subsection{From source-to-target prediction to predictive subspace}

For one target unit and one held-out fold, standard encoding evaluates how accurately a source $X_s$ predicts the target responses $Y$. We report the held-out correlation between predicted and observed target responses as prediction accuracy. In addition to this score, we retain the target-space subspace spanned by the predicted responses.

A ridge-regularized low-rank linear fit \citep{izenman1975rrr,semedo2019communication} induces an orthonormal target basis
\[
    Q_s\in\RR^{q\times k_s}, \qquad Q_s^\top Q_s=I ,
\]
whose columns span the target response patterns predicted from source $s$. The corresponding source-induced predictive subspace is represented by the projector
\[
    P_s = Q_sQ_s^\top .
\]
This projector maps any direction in the target response space onto the predictive subspace induced by source $s$. Because all sources are represented by projectors in the same target response space, brain sources and model sources can be compared despite having different native coordinate systems. Appendix~\ref{app:formal} gives the full indexed algorithm, standardization rules, and hyperparameter selection.

\subsection{Reproducible target reference}

The \emph{reproducible target reference} is defined independently of any source. It identifies target-brain response dimensions that are reproducibly recovered from repeated measurements of the same images. Formally, this reference is an expectation over possible trial partitions. Empirically, we approximate this expectation with random split-half partitions of repeated trials.

For each split $b$, repeated responses to the same held-out images are divided into two averaged target views, $Y_b^{(1)}$ and $Y_b^{(2)}$. We fit target-to-target predictions in both directions, yielding target-side bases $Q_{b,1\to2}$ and $Q_{b,2\to1}$. The reproducible target reference is defined as the average projector
\[
    \TargetRef
    =
    \frac{1}{2N_{\mathrm{split}}}\sum_{b=1}^{N_{\mathrm{split}}}
    \left(
    Q_{b,1\to2}
    Q_{b,1\to2}^{\top}
    +
    Q_{b,2\to1}
    Q_{b,2\to1}^{\top}
    \right).
\]
Its eigendecomposition
\[
    \TargetRef
    =
    U\operatorname{diag}(\omega)U^\top
\]
defines target-reference directions $u_j$ and reference weights $\omega_j$. We use ``direction'' for the mathematical vector $u_j$ and ``dimension'' when emphasizing its interpretation as a response dimension in the target brain. A larger $\omega_j$ indicates that direction $u_j$ is more consistently recovered across trial partitions and therefore receives greater weight in the recovery profile. Unlike PCA, the reproducible target reference is ranked by predictive reproducibility across repeated measurements rather than by response variance. Appendix~\ref{app:split_validity} provides the formal motivation, and Appendix~\ref{app:robustness} provides empirical controls.

\subsection{Recovery profile}

We define the \emph{recovery profile} as the structural summary obtained by comparing a source-induced predictive subspace with the reproducible target reference. It asks how much the prediction overlaps with the reproducible response dimensions of the target brain, and which target-reference dimensions are recovered.

For a source $s$, directional reference coverage measures how much the target-reference direction $u_j$ overlaps with the predictive subspace:
\[
    \DirCov_{s,j}
    =
    \left\|
    Q_s^\top u_j
    \right\|_2^2
    =
    u_j^\top P_s u_j.
\]
This value is $1$ when $u_j$ lies fully within the predictive subspace and $0$ when it is orthogonal to that subspace. Thus, $\DirCov_{s,j}$ measures whether target-reference direction $j$ is recovered from source $s$.

Because individual reference directions can be unstable when nearby reference weights are similar, the main figures emphasize top-$k$ reference coverage,
\[
    \TopKCov_{s,k}
    =
    \frac{
    \sum_{j=1}^{k}\omega_j\DirCov_{s,j}
    }{
    \sum_{j=1}^{k}\omega_j
    },
\]
the reference-weighted overlap between the source-induced predictive subspace and the first $k$ target-reference directions. At each $k$, the curve asks what weighted fraction of the leading reproducible target reference is recovered from the source. This is a prefix average, not a cumulative sum, and it is interpreted alongside prediction accuracy rather than replacing it.

A compact summary of the displayed curve is the profile mean
\[
    \ProfileMean_s^{(K)}
    =
    \frac{1}{K}\sum_{k=1}^{K}\TopKCov_{s,k}.
\]

\subsection{Brain-to-brain reference and model comparison}

Recovery profiles are computed for both brain and model sources using the same target unit and the same reproducible target reference. When the source is another subject's brain responses, the resulting brain-to-brain recovery profile provides a human reference for that target unit. This profile shows which target-reference dimensions are typically recoverable from other human brains.

When the source is a model representation, the resulting model-to-brain recovery profile is interpreted relative to this human reference. This comparison asks whether the model recovers the same target-reference dimensions that are recoverable from other subjects' brains, or whether it achieves similar prediction accuracy while recovering different dimensions in the target response space.

In the full analysis, recovery profiles are computed by outer fold and averaged. All main figures display $K=10$, which captures most reference weight while keeping the profiles readable; human-referenced ratios and full-spectrum diagnostics are defined in Appendix~\ref{app:formal}.

\section{Experimental setting}

All comparisons use the same outer-fold prediction protocol. Repeated target responses on outer-test images define the reproducible target reference; source-to-target fits are trained on outer-training images and evaluated on outer-test images in the same target response space. Brain sources are other subjects' responses, and model sources are fixed vision-model representations.

\paragraph{Targets and folds.}
The main analysis uses the shared-image repeated-response subset of the Natural Scenes Dataset: eight subjects, 515 natural images viewed repeatedly during fMRI, and visual-cortex responses from seven ROIs in each hemisphere (V1v, V1d, V2v, V2d, V3v, V3d, hV4) \citep{allen2022nsd}. Images are divided into five synchronized outer folds. For each target unit and outer fold, repeated target responses on the outer-test images define the reproducible target reference. Source-to-target fits for brain sources and model sources are trained only on the corresponding outer-training images and evaluated on the outer-test images.

\paragraph{Separation of fitting and evaluation.}
Outer-test target responses are used only to define the evaluation reference and to compute held-out prediction accuracy, analogous to using test labels to evaluate a predictor. They are not used to choose source representations, model layers, model states, ranks, regularization parameters, or any other fitting choices. Recovery profiles are therefore held out for source fitting and model selection. They should be interpreted as diagnostics of which target-reference dimensions are recovered from a source, not as additional prediction scores. All comparisons are made within a fixed target unit, outer fold, retained ROI vertices, and target-response standardization; different voxel selections, voxel weightings, or preprocessing choices would define a different evaluation metric.

\paragraph{Sources.}
Brain sources are responses from non-target subjects in the corresponding ROI. Model sources are fixed representations from ResNet-18, ResNet-50 \citep{he2016resnet}, VGG-16 \citep{simonyan2015vgg}, and ViT-B/16 \citep{dosovitskiy2021vit}. Each architecture is evaluated in both pretrained and architecture-matched randomly initialized states. All model layers and representation definitions are fixed before outer-test evaluation. No outer-test target responses are used to choose layers, hyperparameters, or model representations.

\paragraph{Fits and controls.}
Source-to-target prediction uses ridge-regularized low-rank linear fits \citep{izenman1975rrr,semedo2019communication,hoerl1970ridge}. Rank and regularization parameters for predictive subspace fitting are selected by inner cross-validation on the outer-training images. Prediction accuracy is computed with a separately selected ridge readout. Recovery profiles are computed from the source-induced predictive subspaces in the target response space. Profile plots display top-$k$ prefixes through $k=10$; the full reproducible target reference has one target-reference direction per retained target voxel. Controls include matched-rank PCA baselines, random target directions, response permutation, and accuracy-matched pretrained-versus-random pairs.

Appendix~\ref{app:formal} gives the exact folds, standardization rules, representation definitions, hyperparameter grids, and algorithms.

\section{Results}

The empirical analysis asks which response dimensions are recovered when predicting target-brain responses. For each target unit, repeated responses define the reproducible target reference. Each source, whether another subject's brain responses or a model representation, is then evaluated using prediction accuracy and a recovery profile. Prediction accuracy measures aggregate prediction quality, whereas the recovery profile summarizes how strongly the prediction overlaps with the reproducible target reference and which target-reference dimensions are recovered. We summarize this profile using top-$k$ reference coverage, the reference-weighted fraction of leading target-reference dimensions recovered from a source.

\begin{figure*}[t]
    \centering
    \includegraphics[width=\linewidth]{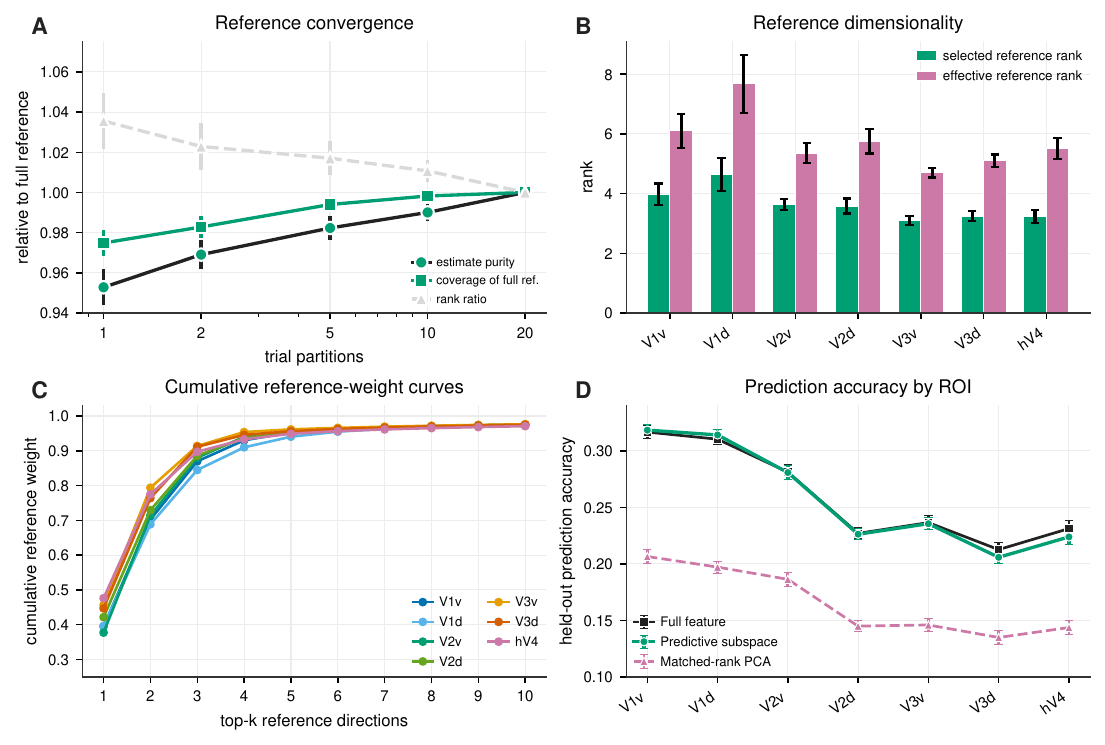}
    \caption{\textbf{Reproducible target references across ROIs.} (A) Independently estimated target references converge toward the estimate obtained from all split-half partitions. (B, C) Reference weights are concentrated but distributed across several target-reference dimensions. (D) Readouts restricted to the selected predictive subspace preserve held-out prediction accuracy relative to full-representation readouts and outperform source-side matched-rank PCA controls. Error bars show SEM; formal metric definitions are in Appendix~\ref{app:formal}.}
    \label{fig:stable_reference}
\end{figure*}

\paragraph{Repeated responses define low-dimensional reproducible target references.}
Figure~\ref{fig:stable_reference} shows that independently estimated target references converge toward the estimate obtained from all split-half partitions, and that the reference weights are concentrated in a small number of target-reference dimensions across visual ROIs. The reference is not a single-dimension effect: the first three dimensions account for 88.6\% of normalized reference weight on average, while the median entropy effective rank is 5.12. Readouts restricted to the selected predictive subspace preserve nearly the same prediction accuracy as full-representation readouts and outperform source-side matched-rank PCA controls. Thus, repeated target-brain responses define a low-dimensional reference that reflects predictive reproducibility across repeated measurements rather than only response variance or rank.

\paragraph{Brain-to-brain recovery profiles provide human references.}
Brain sources recover target-reference dimensions in a structured way. Their top-$k$ reference coverage is high for the leading target-reference dimension but declines as lower-weight dimensions are included. Averaged across target units and brain sources, reference coverage declines from 0.959 at $k=1$ to 0.868 at $k=10$. Thus, brain-to-brain comparison contains more information than a single upper-bound value: it provides a human reference for how strongly target-reference dimensions are typically recovered from another human visual system in each target ROI. Brain-source prediction accuracy is related to the overall height of this curve but does not determine its shape, indicating that brain-to-brain comparison is more informative as a recovery profile than as a scalar ceiling alone. ROI-resolved brain-to-brain profiles are shown in Appendix Figure~\ref{fig:donor_reference}.

\paragraph{Near-equal prediction accuracy can correspond to different recovered dimensions.}\label{sec:matched_result}
When prediction accuracy is nearly fixed, do sources recover the same target-reference dimensions? A case study shows nearly identical prediction accuracy for a pretrained and randomly initialized model source, but the pretrained source has higher top-$k$ reference coverage across the displayed profile. Across source pairs with near-equal prediction accuracy ($|\Delta\mathrm{accuracy}|\le0.01$), pairs of brain sources show small profile differences, whereas source pairs involving models show larger differences. Within-architecture pretrained--random pairs have a positive signed profile-mean difference (mean 0.149; block-bootstrap 95\% CI 0.100--0.199) despite negligible prediction-accuracy difference (mean 0.0024). Using the 90th percentile of profile-mean differences among near-equal-accuracy pairs of brain sources as a descriptive human-variation threshold (0.035), 23 of 29 same-architecture pretrained--random pairs (79\%) exceed this threshold, and 21 of 29 (72\%) do so in favor of the pretrained model. Thus, many pairs that prediction accuracy would treat as effectively tied are separated by their recovery profiles. The key point is not that pretrained models always predict better, but that similar prediction accuracy can correspond to different recovered dimensions in the target brain's response space. Appendix analyses show the same qualitative separation for human-referenced summaries and high-accuracy/profile-shape diagnostics (Figure~\ref{fig:high_predictivity_shape}).

\begin{figure*}[t]
    \centering
    \includegraphics[width=\linewidth]{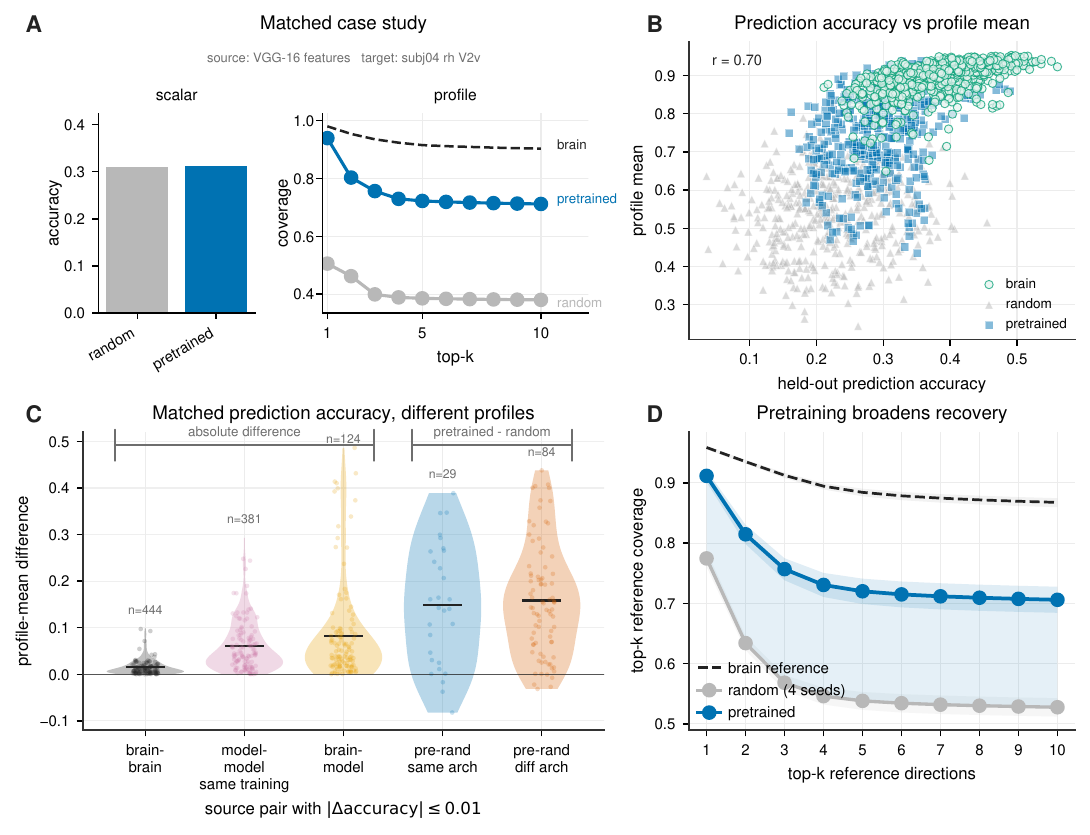}
    \caption{\textbf{Models differ in recovered dimensions beyond prediction accuracy.} (A) A matched VGG-16 case has nearly identical prediction accuracy but different recovery profiles. (B) Prediction accuracy and the overall height of the recovery curve are related but not interchangeable. (C) Near-equal-accuracy pairs show small differences between brain source pairs but larger differences for source pairs involving models; pretrained--random categories use signed pretrained-minus-random differences. (D) Pretraining broadens average top-$k$ reference coverage relative to four-seed random controls, with the brain-to-brain human reference shown as a dashed curve. Shaded bands show 95\% target-unit block-bootstrap intervals.}
    \label{fig:model_recovery}
\end{figure*}

\paragraph{Pretraining broadens recovery of target-reference dimensions.}
The accuracy-matched analysis shows that ImageNet pretraining changes recovery profiles in a way that is not captured by held-out prediction accuracy alone. Panels A--C use random controls initialized with seed 42 for the matched case and pairwise comparisons; in the aggregate profile in Figure~\ref{fig:model_recovery}D, random controls are averaged over four shared initialization seeds. Pretrained model sources exceed the four-seed random mean in profile mean by 0.177 (block-bootstrap 95\% CI: 0.162--0.192; Appendix~\ref{app:random_seed_control}), substantially larger than the overall random-seed standard deviation of 0.027. This pretrained-minus-random gain is not restricted to the leading dimension: the top-$k$ reference coverage difference is 0.137 at $k=1$ and remains between 0.178 and 0.189 for $k=2$--10. Thus, ImageNet pretraining changes which target-reference dimensions the model recovers, not only how accurately it predicts the target responses.

\paragraph{Robustness checks.}
Appendix~\ref{app:robustness} reports trial-partition resampling, fixed-rank variants, PCA and null controls, NSD-synthetic validation, reporting-choice sensitivity, direction-wise held-out prediction controls, random-initialization seed sensitivity, session/run-balanced trial splits, and a disjoint-fold stress test. These analyses preserve the pretrained--random separation, including fixed-rank controls at $k=2$ and $k=3$, show stable profile means under alternative split or fold choices, and confirm that matched-rank target PCA recovers substantially fewer target-reference dimensions than brain-to-brain profiles (0.43 vs.\ 0.85 full-spectrum reference coverage).

\section{Discussion and scope}

\paragraph{Response dimensions make prediction accuracy structurally interpretable.}
This work treats model--brain prediction as a target-space comparison, not only as a held-out accuracy score. Prediction accuracy remains useful, but when it is used as evidence of model--brain alignment, a second question becomes essential: which target-reference dimensions are recovered by the prediction? Brain-to-brain data provide a human reference for this question by showing which target-reference dimensions are typically recoverable from other subjects' brains. The near-equal-accuracy analyses show why this matters: pretrained and randomly initialized models can have similar prediction accuracy while differing in how strongly their predictions overlap with the reproducible target reference and which target-reference dimensions they recover.

\paragraph{Recovery profiles localize recovered and missing target-reference dimensions.}
The recovery profile localizes the model--brain mismatch within the reproducible target reference. A model may recover the leading target-reference dimensions while missing lower-weight dimensions that are typically recovered from other human brains. Alternatively, it may achieve high prediction accuracy with a recovery profile whose shape differs from that of the human brain-to-brain reference. The profile, therefore, makes the evaluation more specific: instead of only reporting that a system predicts well, it shows which target-reference dimensions are recovered and which are missed. Because the profile is organized around a small set of target-reference directions, it also provides a tractable basis for future analyses of what those directions represent.

\paragraph{High recovery is conditional evidence, not whole-model brain-likeness.}
High recovery of the reproducible target reference supports a conditional interpretation of model--brain alignment, rather than a global claim of brain-likeness. It means that, for the evaluated ROI, dataset, preprocessing, and readout class, the source-to-target fit recovers reproducible response dimensions in the target brain's response space. It should not be interpreted as evidence that an entire model representation is brain-like, nor as direction-wise held-out prediction accuracy. This distinction aligns with arguments in the NeuroAI Turing Test that useful neural prediction and stronger claims of brain-likeness should be kept separate \citep{feather2025neuroai}. The same fit also yields source-side predictive coordinates, which can support downstream diagnostics of layer contribution, source-side attribution, and reverse predictivity \citep{muzellec2026reverse}.

\paragraph{The framework complements existing alignment and subspace evaluations.}
The approach is complementary to other structure-aware evaluations. Alignment pattern analysis adds a cross-region relational criterion \citep{hoefling2026onlybrains}, whereas spectral theory decomposes neural prediction through model-side geometry \citep{canatar2023spectral}. Our framework instead fixes a reference in the target brain's response space before evaluating any source. For each target unit, a subject--hemisphere--ROI combination, it asks which target-reference dimensions are recovered from other subjects' brains and from model representations. In this sense, the framework connects ideas from functional alignment and communication subspaces to model--brain evaluation, but uses a reproducible reference defined within the target brain's response space by repeated measurements \citep{chen2015srm,semedo2019communication}.

\paragraph{When to use this framework.}
This framework is useful when the goal is not only to predict a target response but also to understand which parts of the target response space are recovered by different sources. It requires repeated measurements of the target, because those repeats define the reproducible target reference. It is most appropriate for measured response patterns, such as voxel or neural population responses, in which different response dimensions may be recovered from different sources. In this setting, prediction accuracy can be reported alongside a recovery profile: prediction accuracy indicates how well the target responses are predicted, whereas the recovery profile indicates which target-reference dimensions are recovered. The framework is less appropriate when there are no repeated target measurements, or when the target is a single deterministic label rather than a response space with dimensions to recover.

\paragraph{Limitations.}
Empirically, we validate the framework in the early-to-intermediate stages of the human visual cortex. Applications to higher-level cortex, multimodal measurements, or non-neural biological response patterns should re-evaluate the stability of the reproducible target reference and the effects of preprocessing choices. The method also depends on repeated-image data, linear readouts, and low-rank predictive structure. Individual target-reference directions are not interpreted semantically, because predictive subspaces can rotate within nearly equivalent solutions; the main analyses therefore emphasize top-$k$ recovery profiles. Brain-to-brain profiles are empirical human references, not ground-truth definitions of brain-likeness. The matched randomly initialized case study and pairwise analyses use a single fixed initialization seed, whereas the aggregate random profiles in Figure~\ref{fig:model_recovery}D and Appendix Figures~\ref{fig:roi_arch_profiles_core}--\ref{fig:roi_arch_profiles_synthetic} average over four shared initialization seeds; Appendix~\ref{app:random_seed_control} reports seed sensitivity for all main architectures. NSD-core-shared versus NSD-synthetic comparisons test robustness across data regimes rather than a pure stimulus-only manipulation. The split-reference argument assumes that noise specific to one trial view is not systematically reproduced in the other view; session-wise standardization and session/run-balanced split controls reduce, but cannot eliminate, all possible session-correlated effects.

\clearpage
\section*{Acknowledgments}

The first author thanks Takumi Ishimine for helpful discussions related to this work.

\bibliographystyle{plainnat}
\bibliography{references}

\clearpage
\appendix

\section{Experimental setting}
\label{app:setting}

The main text summarizes the information needed to interpret the results. This appendix section defines the concrete scope of the analysis: which dataset, subjects, ROIs, sources, and outputs are included in the primary evaluation. The subsequent formal definition section specifies the exact folds, standardization rules, hyperparameter grids, feature definitions, and metric computations.

\begin{table}[h]
\centering
\caption{\textbf{Experimental setting.} The primary analysis uses the following data, ROIs, sources, and metrics.}
\label{tab:setting}
{\small
\setlength{\tabcolsep}{3pt}
\begin{tabular}{@{}p{0.24\linewidth}p{0.68\linewidth}@{}}
\toprule
Component & Main setting \\
\midrule
Dataset & NSD-core-shared \\
Subjects & 8 subjects; for each target subject, 7 other subjects serve as brain sources \\
Images & 515 shared repeated natural images \\
ROIs & V1v, V1d, V2v, V2d, V3v, V3d, hV4; both hemispheres \\
Outer evaluation & 5 synchronized image folds \\
Target reference & reproducible target reference from split-half target-to-target prediction \\
Model families & ResNet-18/50, VGG-16, ViT-B/16 \\
Model states & pretrained and architecture-matched random controls \\
Main output & recovery profile over target-reference dimensions; brain-source-referenced score as diagnostic \\
\bottomrule
\end{tabular}
}
\end{table}

\paragraph{Dataset selection.}
The primary dataset, NSD-core-shared, is a subset of the Natural Scenes Dataset images that are shared across all eight subjects and have at least three repeated trials for each subject. This rule yields 515 natural images. The robustness dataset, NSD-synthetic, uses the 284 unique synthetic stimuli appearing in the NSD-synthetic experimental ordering for the same eight subjects. In both datasets, image identities are ordered once, and all subjects, hemispheres, ROIs, donor sources, and model sources use that same image order. Five synchronized outer folds are produced from a single fixed random permutation. For NSD-core-shared, each fold has 309 outer-training images and 103 held-out test images, with an additional 103-image chunk reserved but unused in the reported model selection procedure. For NSD-synthetic, the same five-fold rule yields 170--171 outer-training images and 56--57 held-out test images per fold.

\paragraph{Response preprocessing and ROI extraction.}
All fMRI responses are GLMdenoise response-amplitude estimates in fsaverage surface space from NSD and NSD-synthetic \citep{allen2022nsd,gifford2026nsdsynthetic}. The main ROIs are V1v, V1d, V2v, V2d, V3v, V3d, and hV4 in each hemisphere. ROI vertices are selected from the corresponding fsaverage ROI labels; the main analyses do not apply an additional reliability threshold or voxel-selection step beyond these ROI masks. Nonfinite response values are set to zero after standardization, and vertices with nonfinite or near-zero standard deviation are assigned a unit standard deviation before z-scoring.

For NSD-core-shared, trial-level beta estimates are loaded session by session. Within each session and hemisphere, each vertex is z-scored across all trials in that session. Trials corresponding to the 515 selected shared images are then retained. Image-level response matrices used for donor and model source-to-target fits are formed by averaging the session-wise z-scored trial responses across all repeats of each selected image. The repeated-trial target reference uses the same session-wise z-scored trial responses before repeat averaging: for each held-out image, repeated trials are randomly partitioned into two non-empty groups and averaged to form the two sibling views used in the target-reference construction.

For NSD-synthetic, trial beta estimates are available as a single synthetic beta series rather than separate NSD-core sessions. Each vertex is therefore z-scored across all synthetic trials for that subject and hemisphere before any repeat averaging. Image-level response matrices are formed by averaging the z-scored trials assigned to each of the 284 synthetic stimuli. The repeated-trial target reference is constructed with the same image-wise two-view splitting procedure as in NSD-core-shared. Because NSD-core-shared and NSD-synthetic differ in stimulus distribution and response preprocessing, NSD-synthetic is treated as a dataset-shift check rather than a pure replication.

\section{Formal definitions}
\label{app:formal}

This appendix provides the full analysis specification needed to reproduce the quantities in the main text. The central operation is a source-to-target prediction problem: a brain-source response matrix or model-representation matrix is used to predict the voxel responses of one target unit. Prediction accuracy is the mean held-out predictive correlation across target voxels. The structured analysis retains the target-space subspace spanned by the predicted responses and compares that source-induced predictive subspace with the reproducible target reference constructed from repeated target responses.

\paragraph{Terminology overview.}
The appendix follows the same terminology as the main text. The \emph{reproducible target reference} is the set of target-brain response dimensions recovered across repeated trial splits. A \emph{source-induced predictive subspace} is the target-space subspace spanned by responses predicted from one source. A \emph{recovery profile} compares these two objects.

\paragraph{Notational conventions.}
The main text uses \emph{source} for the input side of a source-to-target prediction. In the appendix, we use \emph{brain source} for another subject's brain responses and \emph{model source} for a model representation. When formulas require an index, we use $d$ for a \emph{donor}, meaning a non-target subject used as a brain source for a given target unit. Thus, ``donor'' is only a notational shorthand for brain-to-brain source-to-target prediction, not a separate conceptual object.

\paragraph{Analysis overview.}
For each target unit and outer fold, the analysis proceeds in six steps.
\begin{enumerate}
    \item Source and target matrices are split into outer-training and held-out outer-test images, with standardization statistics estimated only on the relevant training data.
    \item Each brain or model source is fit to the target responses on outer-training images, with rank and subspace regularization selected by inner cross-validation; this produces a source-induced predictive subspace in the target response space.
    \item Held-out repeated target responses are split into two averaged views, and target-to-target prediction between these views defines the reproducible target reference for that fold.
    \item Each source-induced predictive subspace is compared with this reference using directional and top-$k$ reference coverage.
    \item Fold-level curves are averaged to form recovery profiles.
    \item Scalar summaries such as profile mean, brain-source-referenced score, and full-spectrum reference coverage are computed for compact reporting and controls. The recovery profile, not the scalar summary, is the primary object.
\end{enumerate}

\paragraph{Notation.}
We use $s$ for a generic source, $d$ for a non-target subject (donor) used as a brain source, $m$ for a model source, $t$ for the target subject, $r$ for ROI, $f$ for outer fold, $b$ for trial-split repeat, and $j$ for target-reference direction rank. Here, $d$ is only a notational shorthand for a brain source in brain-to-brain prediction. Table~\ref{tab:core_notation} summarizes the main target-space objects and recovery metrics used throughout the appendix.

\begin{table}[h]
\centering
\caption{\textbf{Core notation.} Main target-space objects and recovery metrics used in the appendix.}
\label{tab:core_notation}
{\small
\setlength{\tabcolsep}{3pt}
\begin{tabular}{@{}p{0.30\linewidth}p{0.62\linewidth}@{}}
\toprule
Symbol & Meaning \\
\midrule
$Q_{s\to t,r}^{(f)}$ & target-space basis for the source-induced predictive subspace for source $s$; columns are orthonormal \\
$P_{s\to t,r}^{(f)}$ & orthogonal target-space projector for the source-induced predictive subspace $Q_{s\to t,r}^{(f)}Q_{s\to t,r}^{(f)\top}$ \\
$\TargetRef_{t,r}^{(f)}$ & reproducible target reference; a soft projector in the target response space\\
$u_{t,r,j}^{(f)}, \omega_{t,r,j}^{(f)}$ & target-reference direction and reference weight/eigenvalue from the eigendecomposition of $\TargetRef$; the directions are columns of an orthogonal eigenvector matrix\\
$\DirCov_{s\to t,r,j}^{(f)}$ & directional reference coverage for reference direction $j$ \\
$\TopKCov_{s\to t,r,k}^{(f)}$ & top-$k$ reference coverage \\
$\ProfileMean_{s\to t,r}^{(K)}$ & mean top-$K$ coverage summary \\
$\BrainRefScore_{s\to t,r}^{(K)}$ & brain-source-referenced profile score \\
$\mathrm{ShapeDist}_{s\to t,r}^{(K)}$ & brain-source profile-shape distance \\
$\FullRefCov_{s\to t,r}$ & full-spectrum reference coverage diagnostic \\
\bottomrule
\end{tabular}
}
\end{table}

The reproducible target reference uses the same projection-matrix averaging principle used in consensus subspace estimation \citep{santamaria2016subspaceavg}, with Proposition~2 giving the rank-constrained optimality argument via Ky Fan's principle \citep{fan1951maximum}.

The paragraphs below follow the same order as the analysis: folds and standardization; hyperparameter selection; model representations; predictive fits and scalar readouts; reproducible target reference construction; recovery profiles and scalar summaries. This ordering separates three distinct objects that are easy to conflate: the readout used to report prediction accuracy, the source-induced target basis used for structured recovery, and the reproducible target reference used as the evaluation coordinate system.

\paragraph{Outer folds and standardization.}
Let the 515 shared images be randomly permuted once using a fixed seed, then divided into five synchronized chunks of 103 images each. For outer fold $f$, one chunk is used as the held-out test set, and the next chunk in cyclic order is available as an outer-validation set. The target-reference recovery analysis uses only outer-training and held-out test images; all subspace and readout hyperparameters for this analysis are selected via inner cross-validation on the outer-training images. The auxiliary full-feature versus predictive-subspace correlation comparison in Figure~\ref{fig:stable_reference}(D) uses the outer-validation chunk only to select ridge readout regularization, then reports performance on the held-out test chunk. All z-scoring statistics are estimated only on the training portion of the fit to which they belong and are then applied to validation or held-out images.

\paragraph{Hyperparameter grids and selection.}
There are two regularization parameters in the reported analyses. The subspace regularization $\alpha_{\mathrm{sub}}$ is used when selecting the reduced-rank predictive subspace and when inducing the source-induced target basis. It is selected from
\[
    \{10,10^2,10^3,10^4,10^5,10^6\}.
\]
The readout regularization $\alpha_{\mathrm{read}}$ is used only to fit the low-dimensional ridge readout from selected source coordinates to the target responses, for reporting prediction accuracy. It is selected from
\[
    \{10^{-6},10^{-3},10^{-1},1,10,10^2,10^3,10^4,10^5,10^6\}.
\]
Because the selected readout coordinates and target responses are z-scored and the readout rank is small relative to the number of training images, $\alpha_{\mathrm{read}}=10^{-6}$ is effectively an unregularized least-squares readout in these fits; it is kept as a small positive lower bound for numerical consistency.
The separation is intentional. The subspace regularization defines the predictive geometry retained for target-reference recovery: rank selection, source-side coordinates, and the source-induced target basis. The readout regularization defines only the scalar prediction from those selected coordinates to the full target response. Coupling the two would force the geometry used for recovery profiles and the shrinkage used for scalar prediction to share one tuning parameter, even though they serve different purposes.

For target-reference construction and recovery-profile fits, $(k,\alpha_{\mathrm{sub}})$ is selected by five-fold inner cross-validation over ranks $1\le k\le k_{\max}^{\mathrm{eff}}$, where $k_{\max}^{\mathrm{eff}}=\min(20,p_s,q_r,\lfloor n_{\mathrm{train}}/2\rfloor)$. Each inner split standardizes source and target responses using inner-training statistics only. For each candidate pair, a reduced-rank basis is learned on the inner-training images. Inner-training and inner-validation source responses are projected into the resulting $k$ source coordinates, and a temporary ridge readout from those coordinates to the target responses is fit using the same $\alpha_{\mathrm{sub}}$. The auxiliary prediction-accuracy preservation analysis in Figure~\ref{fig:stable_reference}(D) uses the same selection logic but allows ranks up to 32; this affects only that scalar accuracy comparison, not the target-reference recovery profiles.

The validation score for a candidate pair is the mean voxel-wise Pearson correlation on the inner-validation images. This temporary readout is used only to choose the predictive subspace; it is not the readout used for the held-out predictive-correlation metric. A one-standard-error rule is used: find the best mean validation score, retain all $(k,\alpha_{\mathrm{sub}})$ pairs within one standard error of that best score, choose the smallest rank among those retained pairs, and break remaining ties by higher mean validation score and then smaller $\alpha_{\mathrm{sub}}$. The selected subspace is refit on the full outer-training set. Readout regularization is selected separately via five-fold inner cross-validation, choosing the $\alpha$ with the highest mean validation correlation and breaking ties in favor of the smaller $\alpha$.
In the NSD-core-shared recovery-profile fits, the selected values reflected these different roles. Across 10,640 donor/model source-by-target fold rows, $\alpha_{\mathrm{sub}}$ had median $10^3$ with interquartile range $10^3$--$10^4$, whereas $\alpha_{\mathrm{read}}$ had median $10^{-6}$ with interquartile range $10^{-6}$--$1$. The two selected values were exactly equal in only 0.4\% of rows, and their log-scale correlation was small (Pearson $r=0.044$, Spearman $r=0.076$). Thus, the scalar readout regularizer is empirically not a proxy for the subspace regularizer.

\paragraph{Model representations.}
All model representations are fixed before the held-out evaluation. The main analyses use ImageNet-pretrained, architecture-matched, randomly initialized models with initialization seed 42. For each image, selected activations are averaged over spatial dimensions for convolutional layers or over sequence positions for transformer layers, then concatenated across layers. ResNet-18 uses residual stages 1--4 (960 total features), ResNet-50 uses residual stages 1--4 (3840 features), VGG-16 uses five convolutional feature taps spanning early to final convolutional blocks (1472 features), and ViT-B/16 uses transformer blocks 0, 3, 7, and 11 after the model's standard image preprocessing (3072 features). The same fixed feature definitions are used for pretrained and random states.

\paragraph{Predictive fit and source-induced predictive subspace.}
Given training data $X_s$ and $Y_{t,r}$, each predictive fit solves a ridge-regularized reduced-rank problem
\[
    \min_{B:\operatorname{rank}(B)\le k}
    \|Y_{t,r,z}-X_{s,z}B\|_F^2+\alpha_{\mathrm{sub}}\|B\|_F^2 ,
\]
where $X_{s,z}$ and $Y_{t,r,z}$ are z-scored using training-only statistics. After selecting $(k,\alpha_{\mathrm{sub}})$, the fit is recomputed on the full outer-training set. In standardized coordinates, let
\[
    C=X_{s,z}^{\top}Y_{t,r,z},
    \qquad
    R=X_{s,z}^{\top}X_{s,z}+\alpha_{\mathrm{sub}}I,
    \qquad
    W_{\alpha}=R^{-1}C .
\]
We obtain the source-side predictive directions from the leading eigendirections of $C^{\top}W_{\alpha}$. If
$C^{\top}W_{\alpha}=V\Lambda V^{\top}$, the unnormalized standardized source directions are
\[
    A_{\mathrm{sub}}=W_{\alpha}V_{1:k}.
\]
Their column span is materialized as an orthonormal standardized basis by QR decomposition,
$Q^{X,z}_{s\to t,r,f}=\operatorname{orth}(A_{\mathrm{sub}})$. For feature spaces wider than the number of training images, we use algebraically equivalent dual or smaller eigensystems rather than forming the full source-by-source matrix; the resulting column span is the same reduced-rank ridge source subspace up to numerical roundoff and rotations inside degenerate eigenspaces. Source coordinates for target-basis induction are computed in the same standardized feature space:
\[
    Z^{\mathrm{sub},(f)}_{s\to t,r} = X_{s,z}^{(f)} Q^{X,z}_{s\to t,r,f}.
\]
When an original-feature-coordinate source basis is needed for source-side diagnostics, we store the auxiliary basis
\[
    Q^X_{s\to t,r,f}
    =
    \operatorname{orth}\!\left(
    D_X^{-1}Q^{X,z}_{s\to t,r,f}
    \right),
\]
where $D_X$ is the diagonal matrix of training feature standard deviations. The target-reference recovery metrics do not depend on this auxiliary representation.
The source-induced target basis is the range of the supervised readout from selected source coordinates to target responses. We estimate this readout with ridge regression from $Z^{\mathrm{sub}}$ to the z-scored target responses using the same selected $\alpha_{\mathrm{sub}}$, then take the orthonormal column space of the transposed coefficient matrix:
\[
    Q_{s\to t,r}^{(f)}
    =
    \operatorname{orth}\!\left(
    \hat B_{Z\to Y}^{(f)\top}
    \right).
\]
Here, $\operatorname{orth}(\cdot)$ denotes any orthonormal basis for the column space, so $Q_{s\to t,r}^{(f)\top}Q_{s\to t,r}^{(f)}=I$ and $P_{s\to t,r}^{(f)}=Q_{s\to t,r}^{(f)}Q_{s\to t,r}^{(f)\top}$ is an orthogonal projector. The columns of $Q_{s\to t,r}^{(f)}$ form a target-space basis for the source-induced predictive subspace. The particular basis orientation inside the selected subspace is not meaningful; all coverage metrics depend only on the projector. Using the same $\alpha_{\mathrm{sub}}$ for source-subspace selection and target-basis induction makes the target-side object the target-space range of the regularized predictive solution selected on the source side. A different induction regularization would create an additional, independently tuned target basis and would weaken this source-to-target correspondence. Prediction accuracy is reported separately after fitting a readout with $\alpha_{\mathrm{read}}$.

\paragraph{Readout fitting for prediction accuracy.}
After $Q_{s\to t,r}^{(f)}$ is defined, prediction accuracy is measured by a separate ridge readout from the selected source coordinates to the full target response. The readout uses coordinates
\[
    Z^{\mathrm{read},(f)}_{s\to t,r}=X_{s,z}^{(f)}Q^{X,z}_{s\to t,r,f},
\]
with coordinate and response z-scoring re-estimated inside each readout cross-validation fold. Let $Z_{\mathrm{read},z}^{(f)}$ denote these selected coordinates after the relevant training-fold z-scoring. The fitted readout is
\[
    \hat C_{\mathrm{read}}^{(f)}
    =
    \arg\min_C
    \|Y_{t,r,z}^{(f)}-Z_{\mathrm{read},z}^{(f)}C\|_F^2
    +
    \alpha_{\mathrm{read}}\|C\|_F^2 .
\]
For recovery-profile analyses, the readout parameter $\alpha_{\mathrm{read}}$ is selected by inner cross-validation on the outer-training images using the readout grid above, and the selected readout is refit on the full outer-training set. For the auxiliary full-feature versus predictive-subspace comparison in Figure~\ref{fig:stable_reference}(D), the same grid is selected on the reserved outer-validation chunk. In both cases, prediction accuracy is the mean Pearson correlation across target voxels between the z-scored outer-test responses and the z-scored predictions. It is not computed by reading out only the source-induced target basis, which would evaluate a projected response rather than the full target response. This readout affects prediction accuracy but does not change $Q_{s\to t,r}^{(f)}$, $\TopKCov$, $\ProfileMean$, or $\BrainRefScore$. Thus, prediction accuracy evaluates the prediction of the full target response, whereas the recovery profile evaluates which target-reference dimensions are spanned by the source-induced predictive subspace.

\paragraph{Source-side diagnostics.}
The main evaluation uses the source-induced target basis $Q_{s\to t,r}^{(f)}$, but the same fit also defines a standardized source-side predictive basis $Q^{X,z}_{s\to t,r,f}$ and, when needed, the auxiliary original-coordinate basis $Q^X_{s\to t,r,f}=\operatorname{orth}(D_X^{-1}Q^{X,z}_{s\to t,r,f})$. This basis represents the part of the source representation selected to support the target prediction, not the whole source representation. It induces a source-side projector
\[
    P^X_{s\to t,r,f}
    =
    Q^X_{s\to t,r,f}Q^{X\top}_{s\to t,r,f}.
\]
The selected rank $k$ is the simplest source-side diagnostic for compactness. When nonnegative source predictive strengths $\lambda_{s\to t,r, i}^{(f)}$ are available for the selected coordinates, one may also report an entropy effective rank,
\[
    \exp\!\left(
    -\sum_i \pi_i\log \pi_i
    \right),
    \qquad
    \pi_i=\lambda_i/\sum_\ell \lambda_\ell .
\]
If the source features are concatenated from predefined blocks or layers $\{\mathcal I_\ell\}$, a block-use diagnostic is
\[
    \mathrm{BlockUse}_{s\to t,r,\ell}^{(f)}
    =
    \frac{\|Q^X_{s\to t,r,f}[\mathcal I_\ell,:]\|_F^2}
    {\|Q^X_{s\to t,r,f}\|_F^2}.
\]
Because this quantity depends on feature scaling, preprocessing, and correlations across blocks, it is best interpreted as a descriptive diagnostic or used in conjunction with block-ablation analyses. Fold-to-fold source stability can be measured by normalized projector overlap,
\[
    \mathrm{SourceStability}(f,f')
    =
    \frac{
    \operatorname{tr}\!\left(P^X_{s\to t,r,f}P^X_{s\to t,r,f'}\right)
    }{
    \sqrt{
    \operatorname{tr}\!\left(P^X_{s\to t,r,f}\right)
    \operatorname{tr}\!\left(P^X_{s\to t,r,f'}\right)
    }
    }.
\]
These source-side quantities complement target-reference recovery: recovery profiles identify which target-reference dimensions are recovered in the target brain, whereas source-side diagnostics identify which part of the source representation supports that recovery.

\paragraph{Reproducible target reference.}
For trial-split repeat $b$ among $N_{\mathrm{split}}=20$ repeats, each image's repeated trials are randomly divided into two non-empty groups and averaged to form two paired response views, $Y_b^{(1)}$ and $Y_b^{(2)}$, over the same images. When constructing the evaluation reference for outer fold $f$, this repeated-trial procedure is applied only to the images in that fold's held-out target set.

The two within-target fits induce target bases $Q_{b,1\to2,t,r}^{(f)}$ and $Q_{b,2\to1,t,r}^{(f)}$ using the same $(k,\alpha_{\mathrm{sub}})$ selection rule described above, with the images available to that within-target fit serving as the data pool for inner cross-validation. Their projectors are averaged:
\[
    \TargetRef_{t,r}^{(f)}
    =
    \frac{1}{2N_{\mathrm{split}}}\sum_{b=1}^{N_{\mathrm{split}}}
    \left(
    Q_{b,1\to2,t,r}^{(f)}
    Q_{b,1\to2,t,r}^{(f)\top}
    +
    Q_{b,2\to1,t,r}^{(f)}
    Q_{b,2\to1,t,r}^{(f)\top}
    \right).
\]
This is the reproducible target reference. Its eigendecomposition is
\[
    \TargetRef_{t,r}^{(f)}
    =
    U_{t,r}^{(f)}
    \operatorname{diag}\!\left(\omega_{t,r}^{(f)}\right)
    U_{t,r}^{(f)\top}.
\]
Here, $U_{t,r}^{(f)}$ is an orthogonal eigenvector matrix, $U_{t,r}^{(f)\top}U_{t,r}^{(f)}=I$, because $\TargetRef_{t,r}^{(f)}$ is symmetric. The reference direction $u_{t,r,j}^{(f)}$ is column $j$ of $U_{t,r}^{(f)}$, and $\omega_{t,r,j}^{(f)}$ is the corresponding eigenvalue. The eigenvalues are nonnegative because $\TargetRef$ is an average of projection matrices. If only the leading reference directions are stored or plotted, they are the corresponding orthonormal columns of this full eigenvector matrix. For spectra and weighted profiles, numerical negative eigenvalues are set to zero, and reference weights are normalized as $\tilde\omega_j=\omega_j/\sum_\ell\omega_\ell$; the prefix ratios are unchanged if raw nonnegative $\omega$ values are used.

For the convergence diagnostic in Figure~\ref{fig:stable_reference}(A), each estimate is reduced to a hard consensus subspace. Its rank is
\[
    k_{\mathrm{cons}}
    =
    \operatorname{round}\!\left(
    \frac{1}{2N_{\mathrm{split}}}
    \sum_{b=1}^{N_{\mathrm{split}}}
    \left(k_{b,1\to2}+k_{b,2\to1}\right)
    \right),
\]
clipped to be at least one, where $k_{b,1\to2}$ and $k_{b,2\to1}$ are the cross-validated ranks selected for the two repeated-trial prediction directions. The hard consensus basis consists of the first $k_{\mathrm{cons}}$ columns of $U_{t,r}^{(f)}$, and its projector is denoted $P_{\mathrm{cons}}$. For a lower-partition estimate $P_m$ and the 20-partition estimate $P_{20}$, the plotted diagnostics are
\[
    \mathrm{coverage}_{m\to20}
    =
    \frac{\tr(P_mP_{20})}{\tr(P_{20})},
    \qquad
    \mathrm{purity}_{m\to20}
    =
    \frac{\tr(P_mP_{20})}{\tr(P_m)},
    \qquad
    \mathrm{rank\ ratio}_{m\to20}
    =
    \frac{\tr(P_m)}{\tr(P_{20})}.
\]
Coverage asks how much of the 20-partition consensus subspace is covered by the lower-partition estimate; purity asks how much of the lower-partition estimate lies in the 20-partition consensus subspace; rank ratio compares their dimensions.

\paragraph{Recovery profiles.}
For a donor or model source $s$, directional reference coverage is
\[
    \DirCov_{s\to t,r,j}^{(f)}
    =
    \left\|
    Q_{s\to t,r}^{(f)\top}
    u_{t,r,j}^{(f)}
    \right\|_2^2 .
\]
This is the rank-one specialization of projector overlap. It is between 0 and 1: 0 means the source-induced target basis is orthogonal to that reference direction, and 1 means the reference direction lies inside the source-induced target basis.
The main figures use the reference-weighted top-$k$ coverage profile
\[
    \TopKCov_{s\to t,r,k}^{(f)}
    =
    \frac{
    \sum_{j=1}^{k}\omega_{t,r,j}^{(f)}\DirCov_{s\to t,r,j}^{(f)}
    }{
    \sum_{j=1}^{k}\omega_{t,r,j}^{(f)}
    }.
\]
This is a basis-invariant projector-overlap quantity: it depends on the source-induced target projector and the reproducible target reference directions, rather than on the arbitrary orientation of the basis columns within a selected subspace. Because $\TopKCov$ is a weighted average of $\DirCov$ values with nonnegative reference weights, it also lies between 0 and 1. A value near 1 means that the source-induced predictive subspace overlaps strongly with the reference-weighted leading $k$ target-reference directions; a value near 0 means that those directions are mostly outside the source-induced subspace. This prefix quantity is used rather than axis-by-axis curves in the main figures because individual reference directions can rotate within nearly tied eigenspaces, whereas the leading prefix is more stable and more directly reflects recovery of the leading target-reference dimensions. $\TopKCov$ is a prefix average, not a cumulative sum, so the curve need not increase with $k$: adding a lower-ranked direction can lower the average if that direction is weakly covered. Curves are averaged by within-reference rank, so ``rank 1'' means the leading direction in each fold-specific target reference, not the same vector across subjects or folds.

These are coverage metrics, not direction-wise held-out prediction metrics. For example, $\DirCov_{s\to t,r,j}^{(f)}$ can be high when the induced target subspace contains $u_{t,r,j}^{(f)}$ even if the scalar coordinate $Y u_{t,r,j}^{(f)}$ is predicted weakly on held-out images. Conversely, prediction accuracy can be high because of response variance outside the leading displayed reference prefix. The intended interpretation is therefore joint: prediction accuracy reports the strength of prediction, while the recovery profile reports which reproducible target reference directions are recovered by the predictive subspace.

The full recovery profile is defined for all $q_r$ reference directions of $\TargetRef_{t,r}^{(f)}$, where $q_r$ is the number of retained target voxels for the target subject, hemisphere, and ROI. Thus, the maximum profile length is not fixed globally: in the NSD-core-shared analyses, it ranges from 307 to 1117 directions across target units, with a median of 604.5. For any fixed target fold, however, the reference directions and $q_r$ are identical for all donor and model sources, so their profiles are directly comparable. The source-induced target basis $Q_{s\to t,r}^{(f)}$ can have a much smaller cross-validated rank than $q_r$; recovery at later reference directions is then the squared projection of those directions onto that lower-dimensional source-induced basis.

\paragraph{Scalar summaries.}
Top-$k$ reference coverage is first averaged across outer folds:
\[
    \TopKCov_{s\to t,r,k}
    =
    \frac{1}{|\mathcal F|}
    \sum_{f\in\mathcal F}\TopKCov_{s\to t,r,k}^{(f)} .
\]
The profile-mean summary used in the main figures is
\[
    \ProfileMean_{s\to t,r}^{(K)}
    =
    \frac{1}{K}\sum_{k=1}^{K}\TopKCov_{s\to t,r,k}.
\]
Since it is an average of $\TopKCov$ values, $\ProfileMean$ also lies between 0 and 1. It summarizes the overall height of the displayed profile, not the curve's shape. All main recovery-profile figures use $K=10$, i.e., the first ten eigendirections of each held-out target reference. This is a reporting convention chosen after inspecting the reference spectra, not a hard rank imposed on the subspace fits. The full $\TargetRef$ is computed over all retained target voxels, and $(k,\alpha_{\mathrm{sub}})$ selection is allowed to choose ranks up to $k_{\max}^{\mathrm{eff}}$ as described above.

In the main NSD-core-shared analyses, selected predictive ranks have a median of 3.33 across target units, and entropy effective ranks have a median of 5.12. The first ten reference directions capture 97.4\% of normalized reference weight on average across fold-specific target references, so $K=10$ summarizes nearly the full reproducible target reference while keeping the plotted profiles readable. Analyses that need the entire spectrum use $\FullRefCov$ below.

The brain-source-referenced profile score is
\[
    \BrainRefScore_{s\to t,r}^{(K)}
    =
    \frac{\ProfileMean_{s\to t,r}^{(K)}}
    {\med_{d\neq t}\ProfileMean_{d\to t,r}^{(K)}}.
\]
Thus, a brain-source-referenced profile score of 1 means that a source matches the typical brain source in profile-level recovery for that target subject, hemisphere, and ROI. Values above 1 indicate greater profile-mean coverage than the median brain source for that target unit; they do not imply that the full source representation is more brain-like than a human brain. This ratio is brain-source-referenced rather than a general performance correction: its reference value is the empirical brain-source profile summary for the same target, not an estimate of total response reliability. A full-spectrum reference coverage diagnostic is also computed,
\[
    \FullRefCov_{s\to t,r}
    =
    \frac{1}{|\mathcal F|}
    \sum_{f\in\mathcal F}
    \frac{
    \tr\!\left(P_{s\to t,r}^{(f)}\TargetRef_{t,r}^{(f)}\right)
    }{
    \tr\!\left(\TargetRef_{t,r}^{(f)}\right)
    }.
\]
This full-spectrum reference coverage is a reference-weighted projector-overlap fraction, closely related to standard projection-kernel subspace similarity but normalized by the total target-reference weight. It is the quantity plotted for PCA and null controls in Figure~\ref{fig:controls}(C).
The recovery profile is the primary object; scalar summaries are used only for compact comparisons.

For a separate shape diagnostic, let
\[
    H_{t,r,k}=\med_{d\neq t}\TopKCov_{d\to t,r,k}
\]
be the brain-source-median top-$k$ reference-coverage curve for the same target unit. Each curve is normalized by its own mean,
\[
    \widetilde C_{s\to t,r,k}
    =
    \frac{\TopKCov_{s\to t,r,k}}{\ProfileMean_{s\to t,r}^{(K)}},
    \qquad
    \widetilde H_{t,r,k}
    =
    \frac{H_{t,r,k}}{K^{-1}\sum_{\ell=1}^{K}H_{t,r,\ell}},
\]
and define the brain-source profile-shape distance
\[
    \mathrm{ShapeDist}_{s\to t,r}^{(K)}
    =
    \left[
    \frac{1}{K}
    \sum_{k=1}^{K}
    \left(
    \widetilde C_{s\to t,r,k}
    -
    \widetilde H_{t,r,k}
    \right)^2
    \right]^{1/2}.
\]
Unlike $\BrainRefScore$, this distance removes the overall recovery scale and measures whether the top-$k$ curve has a brain-source-like shape. It is used only as a supplementary diagnostic.

A disjoint-fold reference stress test is also computed. For a source-induced target projector from outer fold $f$, $\TargetRef_{t,r}^{(f)}$ is replaced with $\TargetRef_{t,r}^{(f+1)}$, where $f+1$ is the next cyclic outer fold. This next-fold reference is estimated from a different 103-image chunk and is not used to fit that source projector. The resulting next-fold profile mean tests whether the structured signal is tied to the exact held-out image fold used to define the reference.

\paragraph{Uncertainty estimates.}
All confidence intervals for model comparisons use a target-unit block bootstrap \citep{efron1993bootstrap}. A target unit is a subject--hemisphere--ROI combination. When a target unit is sampled, all model entries, architectures, and states associated with that target unit are resampled together. This preserves the dependence among model comparisons that share the same target responses.

\paragraph{Effective rank.}
If $\lambda_i$ are eigenvalues of the reproducible target reference, its entropy effective rank \citep{roy2007effective} is
\[
    \operatorname{erank}(\TargetRef)
    =
    \exp\left(-\sum_i p_i\log p_i\right),
    \qquad
    p_i=\lambda_i/\sum_j \lambda_j.
\]

\section{Justification of the repeated-trial target reference}
\label{app:split_validity}

This appendix justifies the repeated-trial target reference used as an evaluation coordinate system. The goal is limited: we do not claim to recover a unique ground-truth biological subspace. Instead, we show that the construction provides a principled target-space reference from repeated measurements of the same stimuli.

The argument has three parts. First, under a simple additive repeated-measurement model, the population cross-view covariance contains the repeat-stable signal covariance rather than independent-trial noise. Second, averaging split-specific projectors yields a soft consensus reference whose leading eigenspaces maximize average overlap with the split-specific predictive subspaces, up to non-uniqueness when eigenvalues are tied. Third, finite random trial partitions provide a Monte Carlo estimate of this split-distribution reference, with root-mean-square Frobenius error decreasing as $O(N^{-1/2})$. The empirical controls then test whether this reference behaves as a stable evaluation object in the reported fMRI data.

\paragraph{Proposition 1 (cross-view covariance reflects repeat-stable signal covariance).}
For one target subject, hemisphere, and ROI, suppose two trial-averaged response views for the same images can be written as
\[
    Y^{(1)} = S + E^{(1)},
    \qquad
    Y^{(2)} = S + E^{(2)},
\]
where $S$ is the image-locked repeat-stable response component, and $E^{(1)}$ and $E^{(2)}$ are zero-mean trial-noise terms independent of $S$ and of each other, with finite second moments. Then
\[
    \operatorname{Cov}\!\left(Y^{(1)},Y^{(2)}\right)
    =
    \operatorname{Cov}(S).
\]
Thus, the population cross-view covariance underlying cross-half prediction reflects the repeat-stable signal covariance and contains no contribution from trial noise that is independent across the two views.

\noindent\textbf{Proof.}
Expanding the covariance gives
\[
    \operatorname{Cov}(S+E^{(1)}, S+E^{(2)})
    =
    \operatorname{Cov}(S)
    + \operatorname{Cov}(S,E^{(2)})
    + \operatorname{Cov}(E^{(1)},S)
    + \operatorname{Cov}(E^{(1)},E^{(2)}).
\]
The last three covariance terms vanish by the independence assumptions, together with finite second moments, leaving $\operatorname{Cov}(S)$. Thus, the repeated-trial two-view covariance structure is not an identity reconstruction of the same noisy measurements; it reflects the component that is reproducible across independent trial averages, in the same spirit as classical split-half reliability arguments \citep{spearman1910faulty,brown1910experimental}.

The independence assumption is an idealization. In fMRI, session effects, trial-order effects, and other shared fluctuations can create noise components that are partially reproduced across trial views. The proposition, therefore, justifies the covariance-level intuition behind the split-half target-reference construction; it does not imply that a finite-sample predictive fit is determined only by the repeat-stable signal. The fitted subspace can also depend on the prediction method, regularization, rank selection, finite-sample error, and any noise components shared across views. Empirical trial-partition, PCA, null, and disjoint-fold controls test whether the resulting reference behaves as a stable evaluation object in the reported data.

\paragraph{Proposition 2 (average projector consensus).}
Let $P_1, \ldots, P_M$ be the orthogonal projectors induced by the $M=2N_{\mathrm{split}}$ directional repeated-trial predictive bases for a fixed target fold, and let
\[
    \bar P=\frac{1}{M}\sum_{\ell=1}^{M}P_\ell .
\]
Among all rank-$k$ orthogonal projectors $P$, the average overlap
\[
    J(P)=\frac{1}{M}\sum_{\ell=1}^{M}\tr(PP_\ell)
    =
    \tr(P\bar P)
\]
is maximized by any projector onto a top-$k$ eigenspace of $\bar P$.

\noindent\textbf{Proof.}
This is Ky Fan's maximum principle \citep{fan1951maximum}: for any symmetric matrix $\bar P$, a rank-$k$ projector maximizing $\tr(P\bar P)$ is given by a projector onto a span of $k$ eigenvectors associated with the $k$ largest eigenvalues of $\bar P$. If the $k$th and $(k+1)$th eigenvalues are tied, the maximizer need not be unique. Since $\bar P$ is the mean of the split-specific projectors, its leading eigenspaces provide rank-constrained consensus subspaces that maximize average overlap with the repeated-trial predictive bases.

Propositions 1 and 2 justify the two parts of $\TargetRef$. Proposition 1 explains why using independent repeated-trial views suppresses trial noise that does not reproduce across views. Proposition 2 explains why the eigenspectrum of the averaged projector provides a principled soft reference: large eigenvalues correspond to target directions, or eigenspaces, repeatedly selected by the two-view predictive bases, while unstable directions receive small mass. The empirical construction keeps the full soft projector for recovery profiles rather than committing to a single hard rank.

\paragraph{Proposition 3 (Monte Carlo consistency of the split-reference).}
Fix a target subject, hemisphere, ROI, outer fold, and the observed repeated-trial responses in that fold. Let $\xi_b$ denote the random trial partition used in split repeat $b$, and assume that $\xi_1,\ldots,\xi_N$ are drawn independently from the same partition distribution. For a partition $\xi_b$, let
\[
    P_{b,1\to2}
    =
    Q_{b,1\to2}Q_{b,1\to2}^{\top},
    \qquad
    P_{b,2\to1}
    =
    Q_{b,2\to1}Q_{b,2\to1}^{\top}
\]
be the target-side orthogonal projectors induced by the two repeated-trial predictive fits, and define
\[
    H_b
    =
    \frac{1}{2}
    \left(
    P_{b,1\to2}+P_{b,2\to1}
    \right).
\]
The empirical reproducible target reference with $N$ split repeats is
\[
    \widehat R_N
    =
    \frac{1}{N}\sum_{b=1}^{N}H_b .
\]
Let $R_\star=\mathbb E_{\xi}[H_b\mid\mathcal D]$ be the split-distribution expectation conditional on the observed repeated responses $\mathcal D$. Then
\[
    \mathbb E[\widehat R_N\mid\mathcal D]=R_\star,
\]
and $\widehat R_N\to R_\star$ almost surely and in $L^2$ as $N\to\infty$. Moreover, if every split-specific projector has rank at most $k_{\max}$, then
\[
    \mathbb E\!\left[
    \|\widehat R_N-R_\star\|_F^2
    \,\middle|\,\mathcal D
    \right]
    \le
    \frac{k_{\max}}{N}.
\]
Thus, the mean-squared Frobenius error is $O(N^{-1})$, and the root-mean-square Frobenius error from using a finite number of random trial partitions decays as $O(N^{-1/2})$.

\noindent\textbf{Proof.}
Condition on the observed repeated responses $\mathcal D$. The only remaining randomness is the trial partition $\xi_b$, so the matrices $H_b$ are independent and identically distributed symmetric positive semi-definite random matrices. Since $P_{b,1\to2}$ and $P_{b,2\to1}$ are orthogonal projectors, $0\preceq H_b\preceq I$ and $\|H_b\|_{\mathrm{op}}\le1$. Unbiasedness follows immediately from the linearity of expectation. For $L^2$ convergence, independence gives
\[
    \mathbb E\!\left[
    \|\widehat R_N-R_\star\|_F^2
    \,\middle|\,\mathcal D
    \right]
    =
    \frac{1}{N}
    \mathbb E\!\left[
    \|H_1-R_\star\|_F^2
    \,\middle|\,\mathcal D
    \right].
\]
Because $R_\star=\mathbb E[H_1\mid\mathcal D]$, the variance term is bounded by $\mathbb E[\|H_1\|_F^2\mid\mathcal D]$. The eigenvalues of $H_1$ lie in $[0,1]$, so
\[
    \|H_1\|_F^2
    =
    \tr(H_1^2)
    \le
    \tr(H_1).
\]
For notational simplicity in this bound, write $P_{1\to2}=P_{1,1\to2}$ and $P_{2\to1}=P_{1,2\to1}$. If the two split-specific projectors have rank at most $k_{\max}$, then
\[
    \tr(H_1)
    =
    \frac{1}{2}
    \left[
    \tr(P_{1\to2})+\tr(P_{2\to1})
    \right]
    \le
    k_{\max}.
\]
This proves the stated $L^2$ bound. Almost-sure convergence follows by applying the strong law of large numbers entry-wise to the bounded matrix entries of $H_b$; the target response dimension is finite, so entry-wise convergence implies Frobenius-norm and operator-norm convergence.

Therefore,
\[
    \left(
    \mathbb E\!\left[
    \|\widehat R_N-R_\star\|_F^2
    \,\middle|\,\mathcal D
    \right]
    \right)^{1/2}
    \le
    \sqrt{\frac{k_{\max}}{N}}.
\]
Thus, the mean-squared Frobenius error is $O(N^{-1})$, and the root-mean-square Frobenius error decays as $O(N^{-1/2})$.


\paragraph{Remark on the expectation being estimated.}
The convergence above is conditional on the observed repeated responses. It says that increasing the number of random trial partitions removes Monte Carlo error in the split-reference construction. It does not, by itself, assert that the finite-trial reference equals a unique biological ground-truth subspace. Proposition 1 explains why, under an additive repeated-measure model, the population cross-view covariance underlying cross-half prediction reflects the repeat-stable signal covariance. Stronger population-level consistency of the RRR subspace would require additional assumptions on sample size, regularization, rank selection, and eigengaps.

\section{Robustness and validation}
\label{app:robustness}

\begin{figure*}[h]
    \centering
    \includegraphics[width=\linewidth]{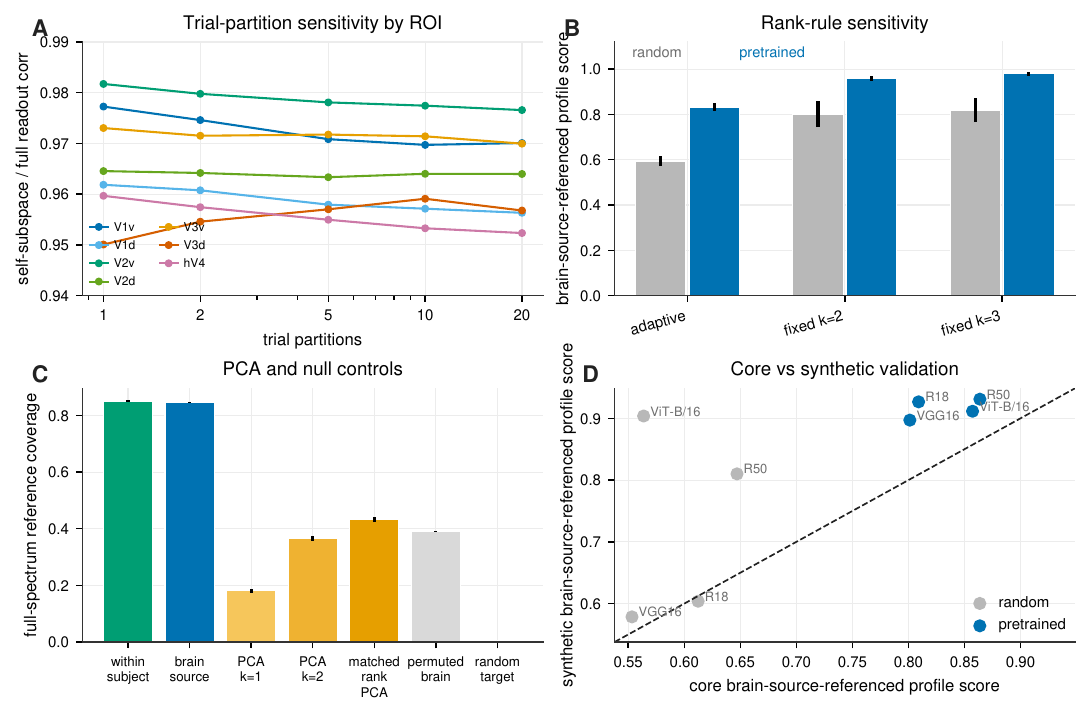}
    \caption{\textbf{Robustness and controls.} (A) Trial-partition resampling sensitivity measured as self-subspace prediction accuracy divided by full-feature readout accuracy. (B) Rank-rule sensitivity for random and pretrained models; error bars show SEM across architectures within each training state and rank rule. (C) Target-side PCA overlaps with part of the reproducible target reference, as expected for reliable high-variance structure, but remains below repeated-trial predictive and brain-to-brain recovery profiles; null controls reduce full-spectrum reference coverage. Error bars show SEM across candidate-by-target entries. (D) NSD-synthetic provides a dataset-shift validation check rather than a controlled stimulus-only replication.}
    \label{fig:controls}
\end{figure*}

The recovery profiles are not explained by an arbitrary low-dimensional structure or response shuffling (Figure~\ref{fig:controls}). Trial-partition resampling yields reproducible target-reference retention, fixed-rank variants preserve the qualitative pretrained-random separation, and random target controls recover few target-reference dimensions. In the rank-rule control, brain-source-referenced profile scores remain higher for pretrained than random models under the adaptive rule (0.833 vs.\ 0.594), fixed rank 2 (0.958 vs.\ 0.801), and fixed rank 3 (0.980 vs.\ 0.819). The PCA controls in Figure~\ref{fig:controls}(C) are target-side controls, separate from the source-side PCA readout in Figure~\ref{fig:stable_reference}(D): they use principal components of the target response covariance, keeping either a fixed number of components or the rank matched to the held-out reproducible-target-reference rank. Target-side PCA captures some reliable variance, but recovers substantially fewer target-reference dimensions than the repeated-trial predictive and brain-to-brain recovery profiles: PCA with one or two components reaches 0.18 and 0.37 full-spectrum reference coverage, respectively, and matched-rank PCA reaches 0.43, compared with 0.85 for brain-to-brain recovery profiles. NSD-synthetic provides an external validation regime with the same qualitative separation. These checks support the interpretation that the framework measures recovery of target-reference dimensions rather than a byproduct of rank, response variance, or one dataset.

We also tested whether the repeated-trial reference depends on random partitions that accidentally align with the NSD acquisition structure. We rebuilt all 560 held-out target references using trial partitions selected to balance the two sibling views in both session and global-run compositions. The resulting references were nearly identical to the main references (soft reference similarity $0.994\pm0.0002$ SEM; hard reference capture $0.980\pm0.003$), and profile means computed against the session/run-balanced references were almost unchanged (overall Pearson $r=0.996$; mean shift $-0.0015$). The pretrained--random profile-mean difference was likewise unchanged (0.2140 with the main reference vs.\ 0.2143 with the session/run-balanced reference), indicating that the main separation is not driven by an obvious session/run partition artifact.

\section{Sensitivity to profile summaries and directional prediction}
\label{app:sensitivity_controls}

The main text reports profile means with $K=10$ and defines near-equal-accuracy pairs using $|\Delta\mathrm{accuracy}|\le0.01$. Table~\ref{tab:sensitivity_controls} shows that the main pretrained--random separation is not driven by these reporting choices. Recomputing the profile mean with $K=5$, $K=10$, or $K=20$ gives nearly identical paired pretrained-minus-random differences, and relaxing the near-equal-accuracy threshold from 0.005 to 0.03 preserves a positive signed difference.

We also computed a direction-wise held-out prediction control. For each saved source-induced basis, we refit only the final ridge readout on the outer-training images, projected held-out predictions and responses onto the stable reference directions, and measured $\mathrm{corr}(\widehat Y u_j, Y u_j)$. This control is deliberately separate from $\DirCov$: it asks whether directions with more extensive geometric coverage are also better predicted on held-out images. Across donor, pretrained, and random sources, $\DirCov$ is positively related to direction-wise held-out prediction, while remaining a distinct subspace-coverage diagnostic. Direction counts in Table~\ref{tab:sensitivity_controls} are descriptive; directions, folds, and sources are not treated as independent inferential samples.

\begin{table*}[h]
\centering
\caption{\textbf{Sensitivity and direction-wise prediction controls.} K sensitivity uses all architecture-matched pretrained--random pairs and reports paired pretrained-minus-random profile-mean differences. Matched-threshold sensitivity uses $K=10$ and varies the allowed absolute difference in prediction accuracy. Direction-wise prediction controls correlate $\DirCov$ with held-out scalar-coordinate prediction along the same reference directions. Direction counts are descriptive, not independent sample counts; confidence intervals, where shown, are target-unit block bootstraps.}
\label{tab:sensitivity_controls}
\scriptsize
\begin{tabular}{@{}llrrrr@{}}
\toprule
Control & Setting & $n$ & Effect & 95\% CI & Additional summary \\
\midrule
$K$ sensitivity & $K=5$ & 448 & $+0.214$ & $[0.199,0.228]$ & \\
 & $K=10$ & 448 & $+0.214$ & $[0.198,0.229]$ & \\
 & $K=20$ & 448 & $+0.213$ & $[0.197,0.228]$ & \\
\midrule
Matched threshold & $|\Delta\mathrm{corr}|\le0.005$ & 17 & $+0.133$ & $[0.070,0.196]$ & 82\% positive \\
 & $|\Delta\mathrm{corr}|\le0.010$ & 29 & $+0.149$ & $[0.103,0.200]$ & 90\% positive \\
 & $|\Delta\mathrm{corr}|\le0.020$ & 60 & $+0.140$ & $[0.111,0.173]$ & 92\% positive \\
 & $|\Delta\mathrm{corr}|\le0.030$ & 94 & $+0.144$ & $[0.118,0.171]$ & 91\% positive \\
\midrule
Directional prediction & donor & 78{,}400 dirs. & $r=0.83$ & -- & high--low quartile $+0.391$ \\
 & pretrained & 44{,}800 dirs. & $r=0.73$ & -- & high--low quartile $+0.230$ \\
 & random & 44{,}800 dirs. & $r=0.63$ & -- & high--low quartile $+0.135$ \\
\bottomrule
\end{tabular}
\end{table*}

\section{Random-initialization seed sensitivity}
\label{app:random_seed_control}

The matched random-initialized case study and pairwise analyses in the main text use a fixed shared initialization seed, while aggregate random profiles in Figure~\ref{fig:model_recovery}D and Appendix Figures~\ref{fig:roi_arch_profiles_core}--\ref{fig:roi_arch_profiles_synthetic} use four-seed means. To test whether the pretrained--random separation depends on initialization, we repeated feature extraction and target-reference recovery for three additional shared-random seeds for all four main architectures, then summarized each seed over the same target units and folds. Table~\ref{tab:random_seed_control} reports the NSD-core-shared seed control; Figure~\ref{fig:roi_arch_profiles_synthetic} uses the analogous NSD-synthetic seed-control table. Seed-to-seed variability is smaller than the pretrained--random profile-mean gap overall. The effect is especially stable for ViT-B/16; ResNet and VGG random controls show greater seed variability, but their seed-averaged profile means remain below those of the corresponding pretrained models.

\begin{table*}[h]
\centering
\caption{\textbf{Random-initialization seed control.} Random controls use four shared initialization seeds (0, 1, 2, and 42) for each architecture. Profile means use the same $K=10$ summary as the main figures and are averaged across subject--hemisphere--ROI target units. The final column reports the random-seed standard deviation divided by the absolute value of the pretrained-minus-random gap.}
\label{tab:random_seed_control}
\scriptsize
\begin{tabular}{@{}lrrrrrr@{}}
\toprule
Architecture & Pretrained profile mean & Random seed mean & Random seed SD & Random min--max & Pretrained--random & SD / gap \\
\midrule
ResNet-18 & 0.727 & 0.623 & 0.059 & 0.551--0.672 & 0.104 & 0.564 \\
ResNet-50 & 0.776 & 0.582 & 0.061 & 0.496--0.636 & 0.194 & 0.316 \\
VGG-16 & 0.720 & 0.568 & 0.053 & 0.498--0.626 & 0.152 & 0.349 \\
ViT-B/16 & 0.770 & 0.513 & 0.005 & 0.507--0.519 & 0.258 & 0.019 \\
\midrule
Overall & 0.748 & 0.571 & 0.027 & -- & 0.177 & 0.151 \\
\bottomrule
\end{tabular}
\end{table*}

\section{Brain-to-brain recovery profiles}

\begin{figure*}[h]
    \centering
    \includegraphics[width=\linewidth]{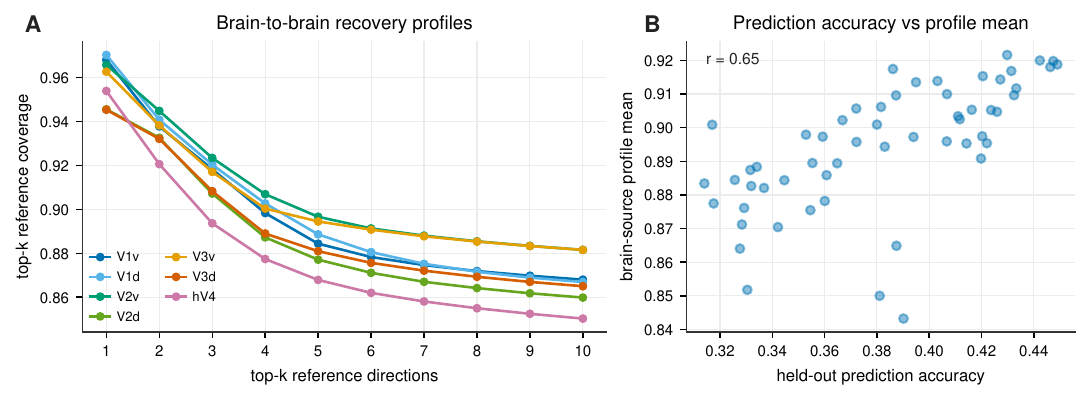}
    \caption{\textbf{Brain-to-brain recovery profiles.} (A) Brain-to-brain top-$k$ reference-coverage curves for all visual ROIs. Top-$k$ reference coverage is a reference-weighted prefix average, not a cumulative sum. (B) Brain-to-brain prediction accuracy and brain-source profile mean are related but not identical, which shows why brain-to-brain comparison is more informative as a recovery profile than a scalar ceiling alone.}
    \label{fig:donor_reference}
\end{figure*}

Figure~\ref{fig:donor_reference} expands the brain-to-brain recovery analysis summarized in the main text. Brain sources provide top-$k$ reference-coverage curves over the same reproducible target reference used for model evaluation. Brain-to-brain prediction accuracy and brain-source profile mean are related but not identical, supporting the use of brain-to-brain data as a recovery profile rather than merely a scalar ceiling.

\section{Dataset-shift analysis}

NSD-synthetic provides an external check using the same subjects and different stimuli \citep{gifford2026nsdsynthetic}. It is used as a robustness analysis rather than as a central result because NSD-core-shared and NSD-synthetic differ not only in stimulus content but also in image selection and response preprocessing.

The synthetic split is therefore interpreted as a dataset-shift stress test rather than a pure replication of the core natural-image setting. Pretrained models retain high brain-source-referenced profile scores across both datasets, whereas random controls are more architecture- and dataset-dependent. In particular, the random ViT shows a high brain-source-referenced profile score on NSD-synthetic, suggesting that the synthetic image set can interact strongly with architectural inductive biases even without learned ImageNet features. This supports interpreting structured recovery in conjunction with the dataset regime, model family, and brain-source profiles, rather than as a standalone ranking score.

\begin{figure*}[h]
    \centering
    \includegraphics[width=\linewidth]{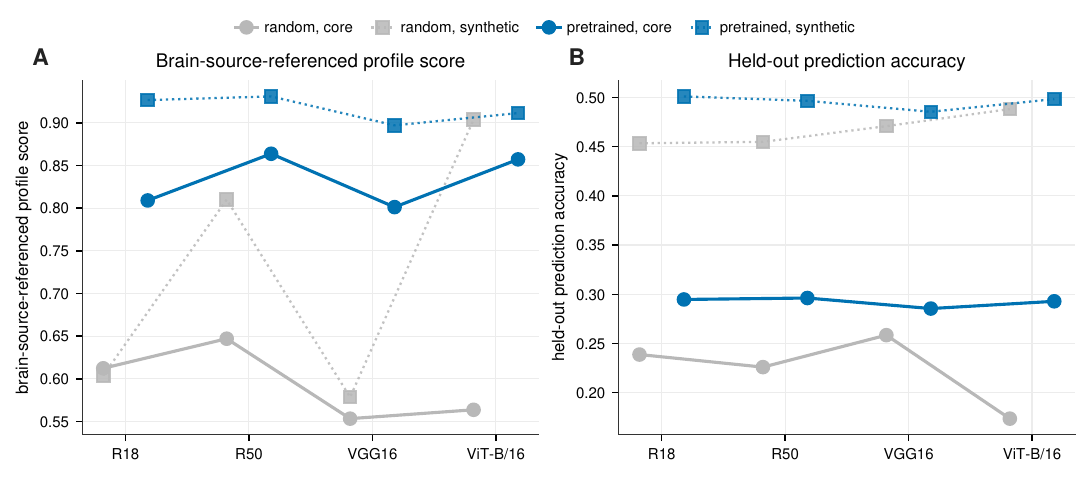}
    \caption{\textbf{Dataset-shift analysis.} NSD-synthetic evaluates structured recovery outside the main natural-image dataset. Pretrained models remain high across datasets, while random controls show stronger architecture dependence, especially for ViT on synthetic images.}
    \label{fig:core_synth}
\end{figure*}

\section{Axis-wise directional coverage}

\begin{figure*}[h]
    \centering
    \includegraphics[width=\linewidth]{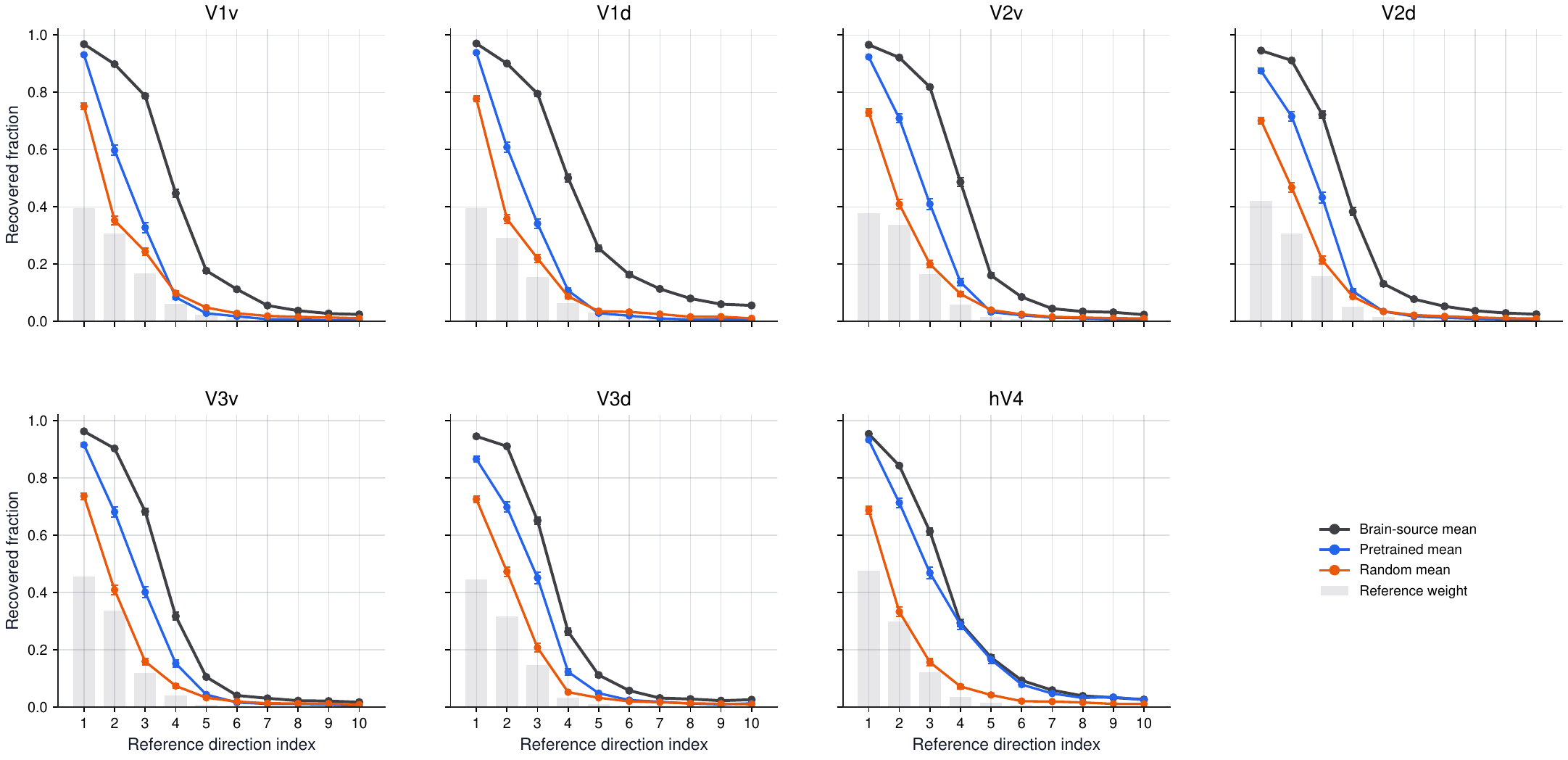}
    \caption{\textbf{Axis-wise directional coverage.} Each panel shows $\DirCov$ for the first ten reference directions in one visual ROI. Lines show brain-to-brain means, pretrained model means, and random-model means; gray bars show the normalized reproducible target reference weight for each direction. Axis-wise coverage is diagnostic, especially for the leading directions, but later low-mass directions are weaker and more variable, so the main figures emphasize top-$k$ reference-coverage profiles.}
    \label{fig:direction_recovery}
\end{figure*}

The main figures use top-$k$ reference coverage because it summarizes recovery of a leading reference prefix and avoids overinterpreting individual axes. Figure~\ref{fig:direction_recovery} shows the underlying axis-wise directional coverage, $\DirCov$, as a diagnostic. The leading directions show the same qualitative ordering as the prefix profiles: brain sources recover the leading target-reference directions most strongly, pretrained models recover more of the leading directions than random controls, and later low-weight directions are weaker and more variable. This supports using axis-wise results as a diagnostic view while keeping top-$k$ profiles as the primary result.

\section{ROI-by-architecture model profiles}

\begin{figure*}[h]
    \centering
    \includegraphics[width=\linewidth]{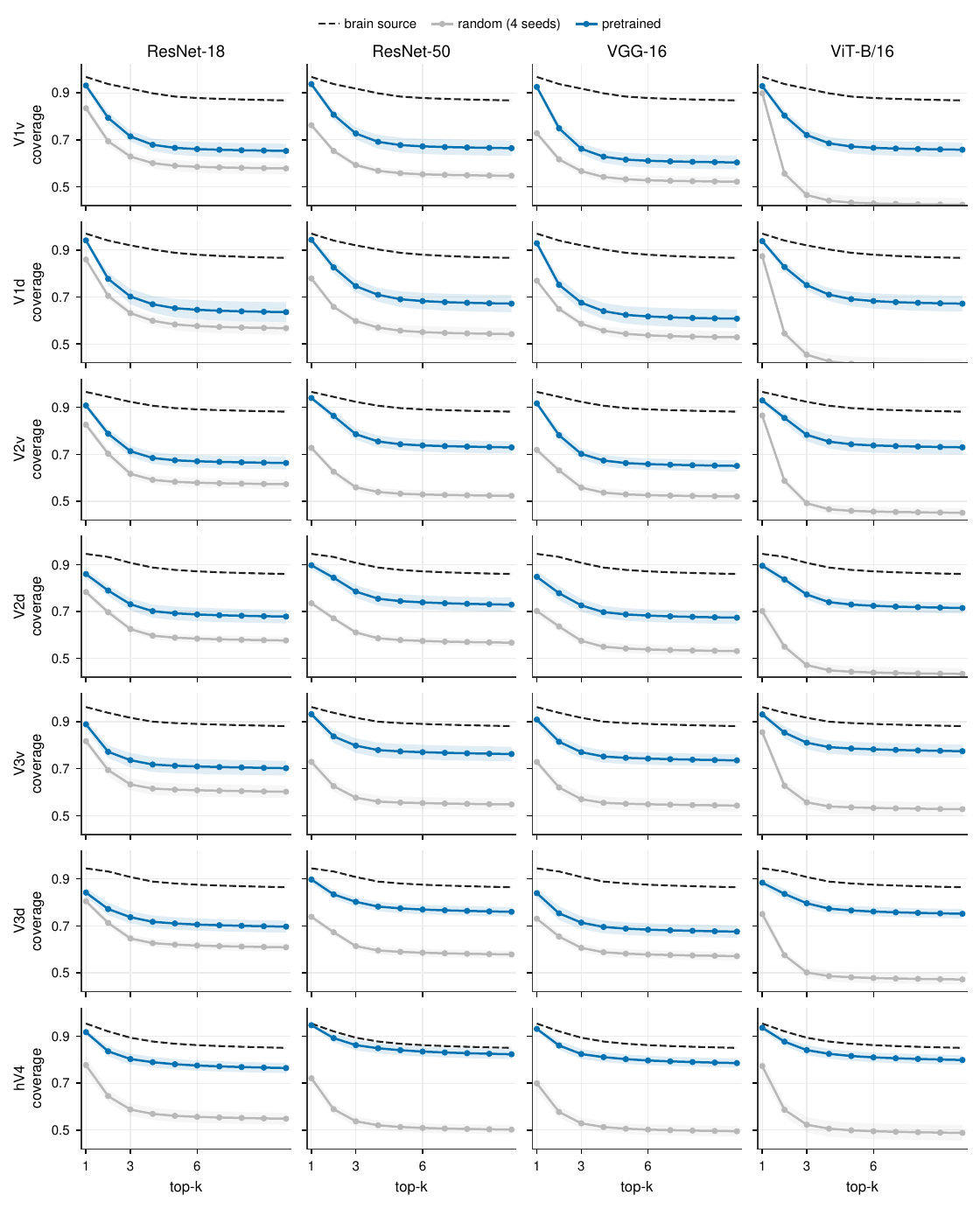}
    \caption{\textbf{NSD-core-shared model recovery profiles by ROI and architecture.} Each panel shows top-$k$ reference coverage for one visual ROI and one architecture. This quantity is a prefix average rather than a cumulative sum. Curves summarize four-seed, randomly initialized models, ImageNet-pretrained models, and the brain-source recovery profile, all over the same reproducible target reference. Shaded bands show SEM across contributing target units after seed averaging for random controls; brain-source bands are computed across brain-source-by-target entries.}
    \label{fig:roi_arch_profiles_core}
\end{figure*}

\begin{figure*}[h]
    \centering
    \includegraphics[width=\linewidth]{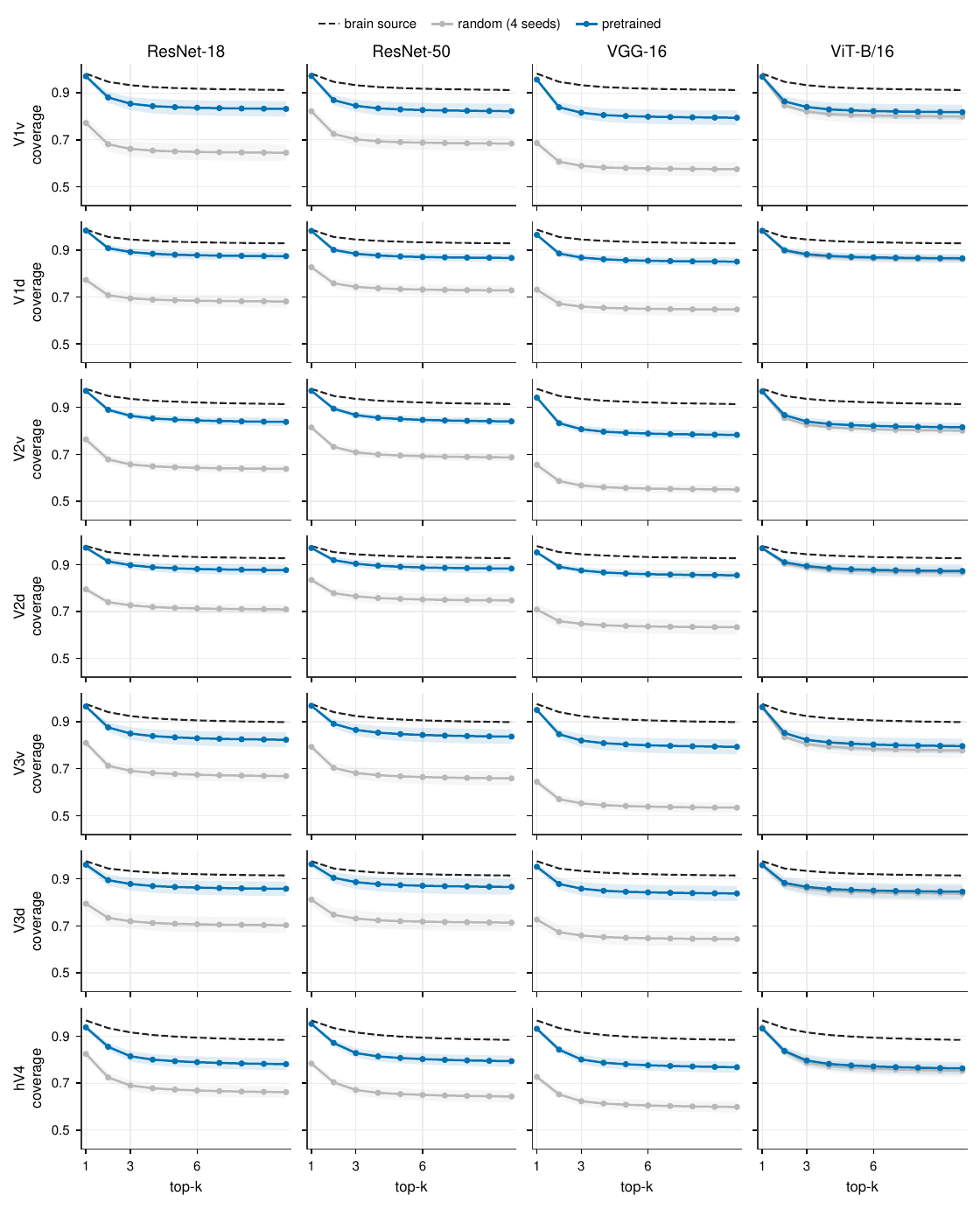}
    \caption{\textbf{NSD-synthetic model recovery profiles by ROI and architecture.} The layout matches Figure~\ref{fig:roi_arch_profiles_core}, but all curves are computed on NSD-synthetic. Top-$k$ reference coverage is a prefix average rather than a cumulative sum. Random controls use the four-seed mean. Shaded bands show SEM across contributing target units after seed averaging for random controls; brain-source bands are computed across brain-source-by-target entries. This independent view highlights dataset-dependent profile changes, including stronger architecture dependence among random controls.}
    \label{fig:roi_arch_profiles_synthetic}
\end{figure*}

The primary figures summarize model behavior in a small number of panels. Figures~\ref{fig:roi_arch_profiles_core} and~\ref{fig:roi_arch_profiles_synthetic} expand the model recovery profiles by visual ROI and architecture for NSD-core-shared and NSD-synthetic separately. Each panel shows the same top-$k$ reference-coverage quantity used in the main figures, averaged across target subjects and hemispheres, with the corresponding brain-source profile shown as a dashed reference curve. This view separates effects that are broad across ROIs from effects that are architecture-, ROI-, or dataset-specific.

\section{Extended model survey}

\begin{figure*}[h]
    \centering
    \includegraphics[width=\linewidth]{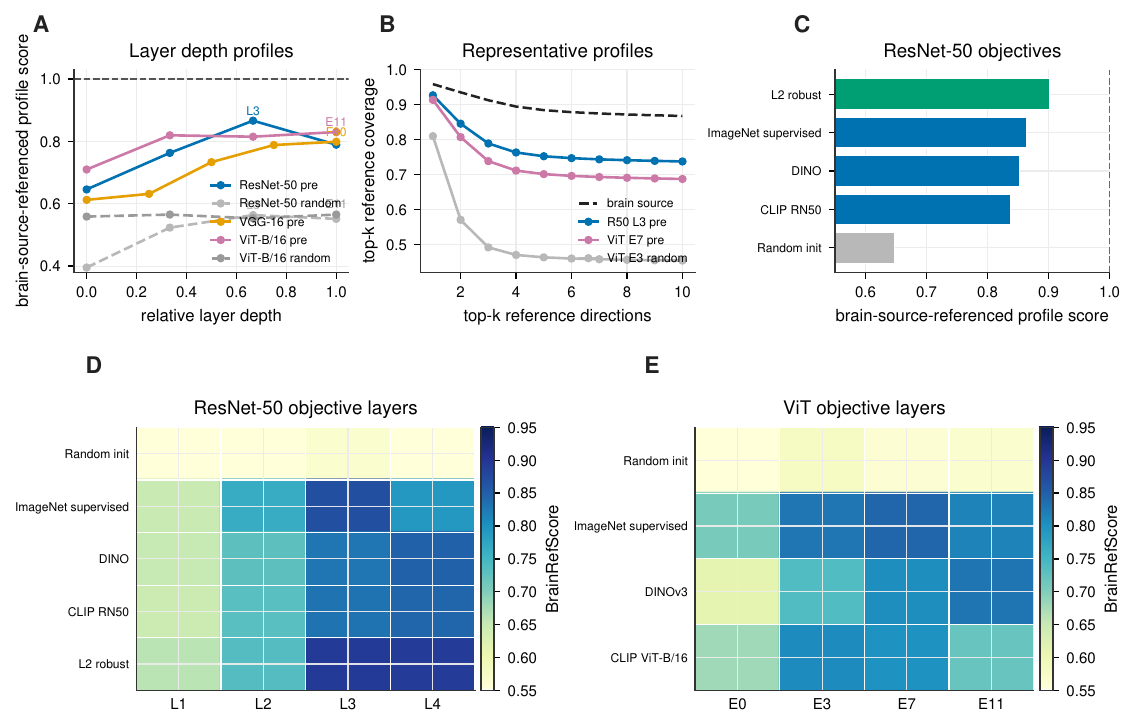}
    \caption{\textbf{Layer-wise and objective diagnostics.} (A) Brain-source-referenced profile score across normalized layer depth for representative random and pretrained models. Labels mark the best layer for each curve. (B) Representative top-$k$ reference-coverage profiles show that high-scoring pretrained layers remain closer to the brain-source profile, whereas random ViT is dominated by the leading direction and drops sharply at later prefixes. (C) ResNet-50 objective sweep summary. (D) ResNet-50 objective-by-layer brain-source-referenced profile scores. (E) ViT objective-by-layer scores. These analyses illustrate the framework's reuse and should not be interpreted as a definitive model ranking.}
    \label{fig:layer_objective_profile_survey}
\end{figure*}

Figure~\ref{fig:layer_objective_profile_survey} and Table~\ref{tab:model_metric_summary} report an extended model survey. These analyses are diagnostic rather than part of the primary result. In the standard model families, brain-source-referenced profile scores, which normalize model recovery by the median brain-to-brain recovery profile for the same target unit, peak in intermediate-to-deep layers, while random ViT profiles remain steep and less brain-source-like. In ResNet-50 objective sweeps, we compare ImageNet supervision, DINO self-supervision \citep{caron2021dino}, CLIP language--image supervision \citep{radford2021clip}, and an L2-robust checkpoint from the MadryLab \texttt{robustness} library \citep{engstrom2019robustness}. The robust objective is included because adversarially robust CNNs have been reported to better match several macaque V1 response properties, including predictivity \citep{kong2022robustv1}. In our sweep, the L2-robust model has the highest brain-source-referenced profile score and the flattest recovery profile among the tested objectives. Table~\ref{tab:model_metric_summary} provides a broader metric summary for the main models, objective sweeps, modern-architecture controls, and follow-up models, including ConvNeXt and Swin backbones \citep{liu2022convnext,liu2021swin}, SigLIP/SigLIP2 image-text encoders \citep{zhai2023siglip,tschannen2025siglip2}, DFN-filtered CLIP models \citep{fang2024dfn}, PE-Core models \citep{bolya2025perceptionencoder}, and DINOv3 \citep{simeoni2025dinov3}. These results illustrate reuse of the evaluation framework across model families; they are reported in the Appendix because the main text focuses on the evaluation protocol rather than a definitive model ranking.

\begin{table*}[h]
\centering
\caption{\textbf{Model metric summary.} The table reports the architecture, pretrained weights or checkpoint family, and evaluated layers or blocks used for each source, organized by training state and architecture family. Prediction accuracy is the scalar readout metric. ProfileMean summarizes top-$k$ reference coverage for $K=10$. BrainRef is the brain-source-referenced score, dividing ProfileMean by the median brain-source ProfileMean for the same target unit. Shape distance compares the mean-normalized model profile to the brain-source-median profile; lower values indicate a more brain-source-like shape. Bold entries indicate the best value in each metric column, with lower values considered better for shape distance.}
\label{tab:model_metric_summary}
{\scriptsize
\setlength{\tabcolsep}{2pt}
\renewcommand{\arraystretch}{1.08}
\begin{tabular}{@{}p{0.15\linewidth}p{0.20\linewidth}p{0.24\linewidth}rrrr@{}}
\toprule
Architecture & Weights / checkpoint & Evaluated layers / blocks & Held-out & Profile & BrainRef & Shape dist. \\
\midrule
\multicolumn{7}{@{}l}{\textit{Pretrained / trained sources}} \\
\multicolumn{7}{@{}l}{\quad ResNet family} \\
ResNet-18 & ImageNet-1K supervised weights, version V1 & residual stages 1,\allowbreak{} 2,\allowbreak{} 3,\allowbreak{} and 4 & 0.295 & 0.727 & 0.809 & 0.072 \\
ResNet-50 & ImageNet-1K supervised weights, version V2 & residual stages 1,\allowbreak{} 2,\allowbreak{} 3,\allowbreak{} and 4 & 0.296 & 0.776 & 0.864 & 0.060 \\
ResNet-50 & DINO self-supervised ResNet-50 checkpoint & residual stages 1,\allowbreak{} 2,\allowbreak{} 3,\allowbreak{} and 4 & 0.295 & 0.765 & 0.852 & 0.064 \\
CLIP RN50 visual encoder & OpenAI CLIP RN50 checkpoint & CLIP ResNet visual stages 1,\allowbreak{} 2,\allowbreak{} 3,\allowbreak{} and 4 & 0.293 & 0.753 & 0.838 & 0.068 \\
ResNet-50 & ImageNet L2-robust checkpoint, epsilon 3.0 & residual stages 1,\allowbreak{} 2,\allowbreak{} 3,\allowbreak{} and 4 & 0.312 & 0.811 & 0.902 & 0.043 \\
\addlinespace[1.5pt]
\multicolumn{7}{@{}l}{\quad Other CNNs} \\
VGG-16 & ImageNet-1K supervised weights, version V1 & VGG convolutional feature taps 4,\allowbreak{} 9,\allowbreak{} 16,\allowbreak{} 23,\allowbreak{} and 30 & 0.285 & 0.720 & 0.801 & 0.077 \\
ConvNeXt-Tiny & ImageNet-1K supervised weights, version V1 & network stages 1,\allowbreak{} 3,\allowbreak{} 5,\allowbreak{} and 7 & 0.281 & 0.716 & 0.797 & 0.086 \\
\addlinespace[1.5pt]
\multicolumn{7}{@{}l}{\quad ViT / transformer family} \\
ViT-B/\allowbreak{}16 & ImageNet-1K supervised weights, version V1 & transformer blocks 0,\allowbreak{} 3,\allowbreak{} 7,\allowbreak{} and 11 & 0.293 & 0.770 & 0.857 & 0.057 \\
CLIP ViT-B/\allowbreak{}16 visual encoder & OpenAI CLIP ViT-B/\allowbreak{}16 checkpoint & transformer blocks 0,\allowbreak{} 3,\allowbreak{} 7,\allowbreak{} and 11 & 0.273 & 0.730 & 0.812 & 0.083 \\
DINOv3 ViT-B/\allowbreak{}16 & DINOv3 public pretrained weights & transformer blocks 0,\allowbreak{} 3,\allowbreak{} 7,\allowbreak{} and 11 & 0.280 & 0.740 & 0.824 & 0.076 \\
Swin-Tiny & ImageNet-1K supervised weights, version V1 & network stages 1,\allowbreak{} 3,\allowbreak{} 5,\allowbreak{} and 7 & 0.289 & 0.731 & 0.814 & 0.077 \\
SigLIP So400m/\allowbreak{}14 & v2 WebLI pretrained weights & transformer blocks 0,\allowbreak{} 3,\allowbreak{} 7,\allowbreak{} and 11 & \textbf{0.320} & \textbf{0.835} & \textbf{0.929} & \textbf{0.037} \\
SigLIP So400m/\allowbreak{}16 & v2 WebLI pretrained weights & transformer blocks 0,\allowbreak{} 3,\allowbreak{} 7,\allowbreak{} and 11 & 0.319 & 0.823 & 0.916 & 0.043 \\
SigLIP So400m/\allowbreak{}16 GAP & v2 WebLI pretrained weights & transformer blocks 0,\allowbreak{} 3,\allowbreak{} 7,\allowbreak{} and 11 & 0.318 & 0.809 & 0.900 & 0.048 \\
SigLIP2 B/\allowbreak{}16 & WebLI pretrained weights & transformer blocks 0,\allowbreak{} 3,\allowbreak{} 7,\allowbreak{} and 11 & 0.293 & 0.775 & 0.862 & 0.059 \\
DFN ViT-L/\allowbreak{}14 & DFN-2B pretrained weights & transformer blocks 0,\allowbreak{} 7,\allowbreak{} 15,\allowbreak{} and 23 & 0.281 & 0.768 & 0.855 & 0.067 \\
ConvStem ViT-B & MCLIP2 DFN-DR2B pretrained weights & transformer blocks 0,\allowbreak{} 3,\allowbreak{} 7,\allowbreak{} and 11 & 0.300 & 0.776 & 0.863 & 0.059 \\
PE-Core-B/\allowbreak{}16-224 & public pretrained weights & transformer blocks 0,\allowbreak{} 3,\allowbreak{} 7,\allowbreak{} and 11 & 0.279 & 0.724 & 0.805 & 0.083 \\
PE-Core-L/\allowbreak{}14-336 & public pretrained weights & transformer blocks 0,\allowbreak{} 3,\allowbreak{} 7,\allowbreak{} and 11 & 0.275 & 0.751 & 0.836 & 0.070 \\
\midrule
\multicolumn{7}{@{}l}{\textit{Untrained controls}} \\
\multicolumn{7}{@{}l}{\quad ResNet family} \\
ResNet-18 & random initialization, seed 42 & residual stages 1,\allowbreak{} 2,\allowbreak{} 3,\allowbreak{} and 4 & 0.239 & 0.551 & 0.612 & 0.095 \\
ResNet-50 & random initialization, seed 42 & residual stages 1,\allowbreak{} 2,\allowbreak{} 3,\allowbreak{} and 4 & 0.226 & 0.582 & 0.647 & 0.090 \\
\addlinespace[1.5pt]
\multicolumn{7}{@{}l}{\quad Other CNNs} \\
VGG-16 & random initialization, seed 42 & VGG convolutional feature taps 4,\allowbreak{} 9,\allowbreak{} 16,\allowbreak{} 23,\allowbreak{} and 30 & 0.258 & 0.498 & 0.553 & 0.090 \\
ConvNeXt-Tiny & random initialization, seed 42 & network stages 1,\allowbreak{} 3,\allowbreak{} 5,\allowbreak{} and 7 & 0.117 & 0.450 & 0.500 & 0.212 \\
\addlinespace[1.5pt]
\multicolumn{7}{@{}l}{\quad ViT / transformer family} \\
ViT-B/\allowbreak{}16 & random initialization, seed 42 & transformer blocks 0,\allowbreak{} 3,\allowbreak{} 7,\allowbreak{} and 11 & 0.174 & 0.512 & 0.570 & 0.204 \\
Swin-Tiny & random initialization, seed 42 & network stages 1,\allowbreak{} 3,\allowbreak{} 5,\allowbreak{} and 7 & 0.177 & 0.559 & 0.622 & 0.166 \\
\bottomrule
\end{tabular}

}
\end{table*}

\section{Source-side explanation diagnostics}

\begin{figure*}[h]
    \centering
    \includegraphics[width=\linewidth]{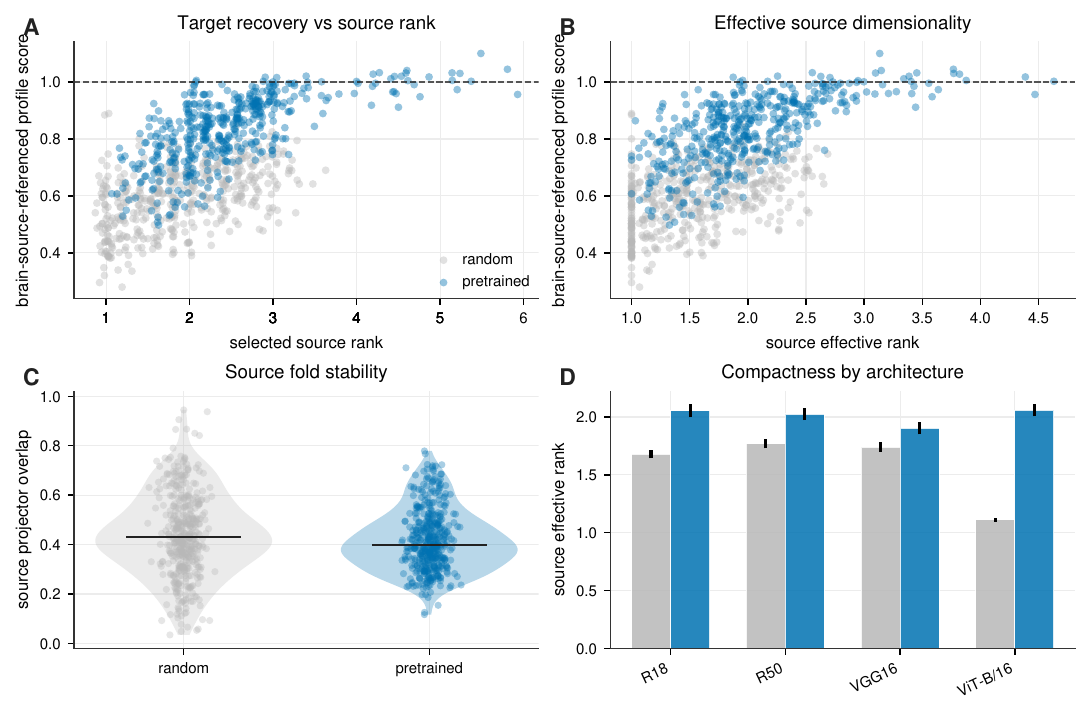}
    \caption{\textbf{Source-side explanation diagnostics.} (A) Brain-source-referenced profile score plotted against the selected source rank. (B) The source effective rank, computed from nonnegative predictive strengths across the selected source coordinates, serves as a diagnostic of compactness. (C) Source-side fold stability is the median normalized projector overlap between source-side predictive subspaces selected on different outer folds for the same model, target subject, hemisphere, and ROI; black horizontal segments mark group medians. (D) Source effective rank summarized by architecture and training state; error bars show SEM across source-by-target entries. These diagnostics identify which part of the source representation supports target-reference coverage; they are not whole-model brain-likeness scores.}
    \label{fig:source_side_diagnostics}
\end{figure*}

The main results identify which target-reference dimensions a model recovers. The same reduced-rank fits also define a source-side predictive subspace that identifies which part of the source representation supports the recovery of target-reference dimensions. Figure~\ref{fig:source_side_diagnostics} illustrates this second-stage diagnostic for the main model set. Selected source subspaces are compact: across model-by-target units, the median selected source rank is 2.2, and the median source effective rank is 1.76. Pretrained models use, on average, slightly higher-dimensional source subspaces than random controls (mean effective rank: 2.01 vs. 1.57). Fold-to-fold source-subspace stability is moderate for both training states (median projector overlap 0.40 for pretrained and 0.43 for random controls), indicating that source-side explanations should be interpreted as diagnostics of the selected predictive subspace rather than as whole-model brain-likeness scores.

\section{High-accuracy and profile-shape checks}

\begin{figure*}[h]
    \centering
    \includegraphics[width=\linewidth]{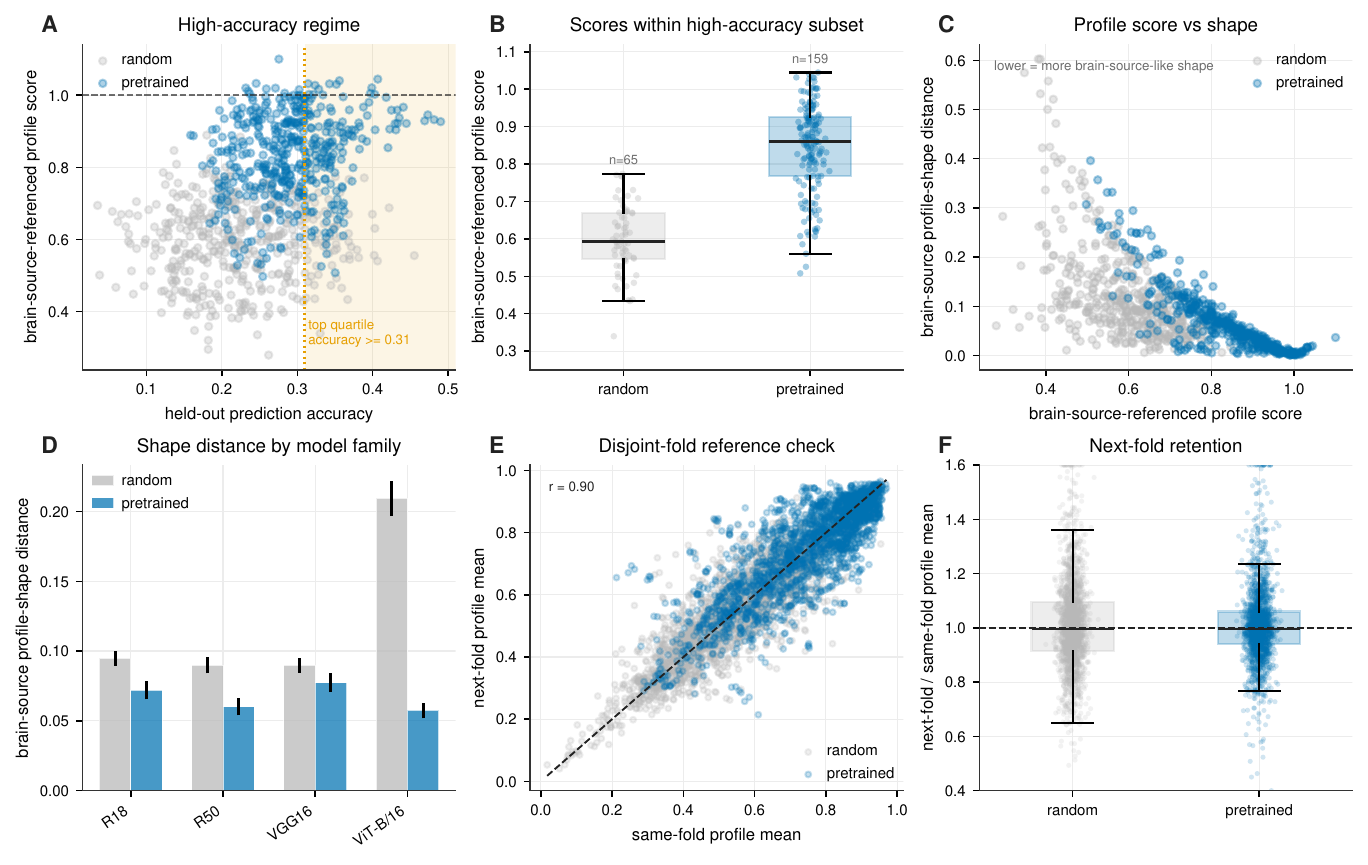}
    \caption{\textbf{High-accuracy and brain-source profile-shape diagnostics.} (A) The top quartile of prediction accuracy is shaded. (B) Within this high-accuracy subset, brain-source-referenced profile scores remain separated between random and pretrained sources; boxplots show medians, interquartile ranges, and 1.5-IQR whiskers. (C) Brain-source profile-shape distance compares the mean-normalized top-$k$ curve with the brain-source-median curve; lower values indicate a more brain-source-like profile shape after removing the overall recovery scale. (D) Shape distance varies by architecture and training state, with random ViT showing the largest deviation; error bars show SEM across source-by-target entries. (E) A disjoint-fold stress test compares the same-fold profile means with profile means computed against the next outer fold's reproducible target reference. (F) Next-fold retention summarizes the ratio between next-fold and same-fold profile means; boxplots use the same convention as in panel B.}
    \label{fig:high_predictivity_shape}
\end{figure*}

Figure~\ref{fig:high_predictivity_shape} evaluates two possible concerns. First, scalar scores can be compressed in a high-accuracy regime, especially in early visual ROIs. The analysis identifies the upper quartile of model-by-target prediction accuracy and re-examines brain-source-referenced profile scores within that subset. The scores remain spread out, indicating that high prediction accuracy does not imply a unique recovery profile. Second, a high brain-source-referenced profile score need not by itself imply a brain-source-like curve shape. The analysis, therefore, reports the brain-source profile-shape distance defined above, which normalizes out the overall profile mean before comparing the shapes of the curves. Finally, the disjoint-fold reference check evaluates each source-induced projector against a reproducible target reference estimated from the next outer fold. Same-fold and next-fold profile means remain strongly correlated ($r=0.90$; mean next/same ratio 1.01), supporting the interpretation that the profile signal is not tied to the exact held-out image fold used to construct the reference.

\section{Data, code availability, and responsible use}
\label{app:code_assets}

The analysis uses NSD, NSD-synthetic, GLMsingle response estimates, ImageNet-pretrained and randomly initialized torchvision/timm model definitions, and public checkpoints used in the appendix diagnostics. Raw NSD-family data, ImageNet images, large intermediate feature or response arrays, and third-party model checkpoints are not redistributed; users must obtain these assets from the original providers under the corresponding access terms and provide local paths in the reproduction workflow.

The analysis code and reproduction scripts are being prepared for public release. This version of the preprint does not yet include a repository URL; an updated version will link the public repository when it becomes available. Until then, the manuscript specifies the data selection, preprocessing, fold construction, model representations, hyperparameter grids, and metric definitions needed to interpret and independently reproduce the reported analyses.

This work analyzes de-identified, previously released human fMRI datasets and collects no new human-subject data. The original NSD and NSD-synthetic publications describe their participant consent and ethics-review procedures. The intended use is evaluation and interpretation of the model--brain alignment, with potential benefits for diagnosing where models match or miss target-reference dimensions. The framework is not a clinical diagnostic, biometric identification method, or deployed decision system. Potential risks are mainly interpretational: high recovery scores should not be read as evidence that an entire model is brain-like, nor should ROI-level fMRI analyses be overextended to individual-level clinical or behavioral claims.

\end{document}